\documentclass[twocolumn,tighten,times]{aastex62}

\usepackage{longtable}

\usepackage{color}

\newcommand{\sn}{$S_N$}
\newcommand{\snz}{$S_{N,z}$}
\newcommand{\fRed}{$f_{Red}$}
\newcommand{\SgmEnv}{$\Sigma_{15}$}
\newcommand{\Mz}{$M_z$}
\newcommand{\re}{$R_e$}
\newcommand{\afe}{[$\alpha$/Fe]}
\newcommand{\LmL}{$L_{max}/L$}
\newcommand{\sersic}{S\'ersic}

\accepted{for publication in The Astrophysical Journal}

\begin{document}

\title{ The ACS Fornax Cluster Survey. III. Globular Cluster Specific Frequencies of Early-Type Galaxies }

\correspondingauthor{Yiqing Liu, Eric W. Peng}
\email{yiqing.liu@physics.ox.ac.uk, halfxianr@gmail.com; peng@pku.edu.cn}
\author{Yiqing Liu}
\affiliation{Sub-department of Astrophysics,
Department of Physics, University of Oxford, Denys Wilkinson Building, Keble Road, Oxford OX1 3RH, United Kingdom}
\author{Eric W. Peng}
\affiliation{Department of Astronomy, Peking University, 
Beijing 100871, China}
\affiliation{Kavli Institute for Astronomy and Astrophysics, Peking University, 
Beijing 100871, China}
\author{Andr\'es Jord\'an}
\affiliation{Instituto de Astrof\'isica, Facultad de F\'isica, Pontificia Universidad Cat\'olica de Chile, 
Av. Vicu\~{n}a Mackenna 4860, 7820436 Macul, Santiago, Chile}
\author{John P. Blakeslee}
\affiliation{National Research Council of Canada, Herzberg Astronomy and Astrophysics Research Centre, 
5071 W. Saanich Road, Victoria, BC V9E 2E7, Canada}
\affiliation{Gemini Observatory, Casilla 603, La Serena, Chile}
\author{Patrick C{\^o}t{\'e}}
\affiliation{National Research Council of Canada, Herzberg Astronomy and Astrophysics Research Centre, 
5071 W. Saanich Road, Victoria, BC V9E 2E7, Canada}
\author{Laura Ferrarese}
\affiliation{National Research Council of Canada, Herzberg Astronomy and Astrophysics Research Centre, 
5071 W. Saanich Road, Victoria, BC V9E 2E7, Canada}
\author{Thomas H. Puzia}
\affiliation{Instituto de Astrof\'isica, Facultad de F\'isica, Pontificia Universidad Cat\'olica de Chile, 
Av. Vicu\~{n}a Mackenna 4860, 7820436 Macul, Santiago, Chile}

% ====================

\begin{abstract}
The globular cluster (GC) specific frequency ($S_N$), defined as the number of GCs per unit galactic luminosity, represents the efficiency of GC formation (and survival) compared to field stars. Despite the naive expectation that star cluster formation should scale directly with star formation, this efficiency varies widely across galaxies. To explore this variation we measure the $z$-band GC specific frequency (\snz) for 43 early-type galaxies (ETGs) from the {\it Hubble Space Telescope (HST)}/Advanced Camera for Surveys (ACS) Fornax Cluster Survey. Combined with the homogenous measurements of \snz\ in 100 ETGs from the HST/ACS Virgo Cluster Survey from \citet{P08}, we investigate the dependence of \snz\ on mass and environment over a range of galaxy properties. 
We find that \snz\ behaves similarly in the two galaxy clusters, despite the clusters' order-of-magnitude difference in mass density. The \snz\ is low in intermediate-mass ETGs ($-20<M_z<-23$), and increases with galaxy luminosity. It is elevated at low masses, on average, but with a large scatter driven by galaxies in dense environments. 
The densest environments with the strongest tidal forces appear to strip the GC systems of low-mass galaxies. However, in low-mass galaxies that are not in strong tidal fields, denser environments correlate with enhanced GC formation efficiencies. 
Normalizing by inferred halo masses, the GC mass fraction, $\eta=(3.36\pm0.2)\times10^{-5}$, is constant for ETGs with stellar masses $\mathcal{M}_\star \lesssim 3\times10^{10}M_\odot$, in agreement with previous studies. 
The lack of correlation between the fraction of GCs and the nuclear light implies only a weak link between the infall of GCs and the formation of nuclei. 
\end{abstract}


% =====================

\section{Introduction}
\label{sec:intro}

Globular clusters (GCs) are old, massive, compact star clusters whose masses range from 10$^4$ to 10$^6~M_\odot$, with a typical size of $\sim$3~pc. They are mostly associated with the old spheroid components of galaxies, with the oldest GCs having ages comparable to the universe.  

Although the exact formation scenario for GCs is still unknown, it is believed that they form in high-pressure and high-density regions with high star formation efficiencies on short timescales. In such environments they might gain enough mass before feedback processes halt star formation \citep[e.g.][]{Elmegreen_Efremov_97,Kruijssen_12}.  One possibility is that GCs are the merger remnants of protogalactic disk fragments in the minihalos at high redshift. \citet{Kimm_16} simulated this scenario in a cosmological context and found that it reproduced well the basic GC properties of mass, size and star formation history. 

In the local universe, young massive star clusters (YMCs) form in regions with high star formation rate (SFR) surface densities \citep{Larsen_Richtler_00III}, and evidence shows that they might be the progenitors of GCs \citep[e.g.,][]{Whitmore_99}. It is likely that GCs also form during violent star formation events, which could happen at high redshift or in gas-rich galactic mergers. 
However they are formed, studying GC systems provides a unique window onto the earliest and most intense episodes of galactic star formation. We can also learn about the subsequent galactic mass assembly histories, in which a majority of GCs in massive galaxies are likely accreted from less massive galaxies in a hierarchical fashion \citep{Cote_98,Tonini_13}. At the same time, such studies provide information about the formation mechanisms of massive star clusters.

While the existence of GC systems beyond the Local Group have has been known for many decades \citep{Baum_55}, the advent of CCDs and the {\it Hubble Space Telescope (HST)} led to the first systematic studies of extragalactic GC systems---their numbers, sizes, luminosities, colors, and stellar populations---as a function of host galaxy properties \citep[e.g.,][]{Larsen_01,P08,Georgiev_10,Jordan_05,Masters_10,Peng_06,Jordan_07a}. Review papers by \citet{Harris_91} and \citet{Brodie_06} have tracked the progress of research in the field of extragalactic GC systems and summarized the state of the art at their respective times.

The number of GCs in a galaxy one of the most basic observables of a GC system. While the number of GCs broadly correlates with galaxy stellar mass, as one might expect, the ratio of the number of GCs to the stellar luminosity is famously not constant. This ratio is widely parameterized as the GC specific frequency (\sn)---defined as the number of GCs per unit of the galactic V-band luminosity, and normalized to $M_V=-15$ \citep{Harris_vdB_81}--- and represents the efficiency of GC formation (and survival) with respect to field stars. 

Previous studies have measured \sn\ in galaxies with a wide range of properties.  
One of the largest individual such studies, and the one most relevant to this paper, is that of \citet{P08}, which used the data from the ACS Virgo Cluster Survey (ACSVCS; \citealt{Cote_04,Jordan_09}) to study \sn\ in 100 Virgo cluster early-type galaxies (ETGs) over a wide range of galactic mass. The Virgo cluster ETGs were found to have a ``U-shaped" relation between \sn\ and galactic luminosity, a shape that they showed mimicked the effects of the stellar mass-halo mass relation of galaxies determined from halo abundance matching. It was found that \sn, or GC mass fractions, were found to be universally low in intermediate-mass galaxies ($-20.5<M_V<-18$), but with upturns occurring at both high and low mass. Other studies using independent data sets show similar results for ETGs (e.g. \citealt{Zaritsky_15}). 

Because GCs do not form in a constant fraction out of field stars, GC formation might be governed by more fundamental regulators. 
\citet{Blakeslee_97a,Blakeslee_99} and \citet{Blakeslee_97b} found that the number of GCs of the bright central galaxies (BCGs) scaled with the total mass of entire galaxy clusters. Furthermore, \citet{McLaughlin_99a} found that in massive ETGs, the data are consistent with a universal fraction of the baryonic mass ending up in GCs.  Later studies show correlations between GC number and the halo mass of individual galaxies \citep{Spitler_Forbes_09,Georgiev_10,Harris_13,Hudson_14}. This is also supported by simulations \citep{Kravtsov_Gnedin_05}. 
These works indicate that the relation between \sn\ and galactic luminosity might come from the similarly shaped distribution of the galactic halo mass-to-light ratio ($M_h/L$) with galactic mass \citep{Wolf_10}. 

Traditionally, studies of \sn\ focus on its relationship with galactic masses or galactic types. Recently, however, the role of environment has turned out to be important, especially at low masses.  
\citet{P08} showed that while \sn\ is systematically higher for both high- and low-mass ETGs, it has large scatter in the low-mass range. \citet{P08} studied this scatter in low-mass ETG \sn, finding that galaxies closer to the cluster center have higher \sn, and suggesting that their \sn\ are regulated by their environment. 

There are two possible (and nonexclusive) explanations for such an environmental dependence. One is that the low-mass galaxies in denser regions had stronger star formation at an early time, and more GCs formed along with it. Alternatively, the star formation of low-mass galaxies stopped earlier in denser environments, and because most GCs formed at very early times, the number of GCs is hardly affected. The decrease of the number of field stars, however, drives up \sn\ in dense environments. 
Both scenarios are supported by the higher \afe\ at smaller clustercentric distances from observations \citep{Liu_16a}. 

Nonetheless, more evidence is needed to expand our understanding of this apparent environmental dependence of \sn. The correlation shown in \citet{P08} is driven by the high \sn\ of the innermost galaxies, and no clear dependence is found in the lower-mass sample of \citet{Miller_Lotz_07}. 
The environmental dependence of \afe\ is similarly tantalizing. \citet{Liu_16c} found that only the densest environments affect the \afe\ distribution of low-mass ETGs. 

Therefore, the importance of environment for GC formation is a question. 
A place to study environmental effects is galaxy clusters, as the mass density changes with clustercentric distance.  In addition, comparing galaxy clusters with different densities provides further constraints. 
\citet{P08} studied the \sn\ of ETGs and their relations with the masses and environments in the nearest galaxy cluster, and we extend this work by adding samples from the next-nearest galaxy cluster. 

As a complementary program to the ACSVCS, the ACS Fornax Cluster Survey \citep[ACSFCS;][]{Jordan_07b} imaged 43 galaxies in the Fornax Cluster, a galaxy cluster that, compared to Virgo, is $\sim\!\!20\%$ more distant \citep{Blakeslee_09}, an order of magnitude less massive, and dynamically older. Because they have similar virial radii, the mass density of Fornax is significantly lower. 
In this work, we measure the \sn\ of 43 ETGs from the ACSFCS sample. Combining with the results of 100 ETGs from the ACSVCS sample, which were measured by \citet{P08}, we study the dependence of \sn\ on mass and environment among ETGs with a significant sample size. 

The paper is structured as follows. Our data are introduced in $\S$~\ref{data}. $\S$~\ref{count} describes our measurement of the number of GCs.  We then present the relationships between \sn, galactic masses, and environments in $\S$~\ref{result}, and present a related discussion in $\S$~\ref{discussion}. In $\S$~\ref{summary}, we list our main conclusions. \\

% ======================

\section{Observations and Data}
\label{data}

The ACSFCS program imaged 43 ETGs in the Fornax Cluster with the Hubble Space Telescope (HST) Advanced Camera for Surveys \citep[ACS;][]{Ford_98}. 
This is a complete sample of the Fornax galaxies that are brighter than $B_T \sim 15.5$ ($M_B \sim -16$) mag, covering the morphological types of E, S0, SB0, dE, dE,N, dS0, and dS0,N. It includes 41 galaxies from the Fornax Cluster Catalog \citep[FCC;][]{Ferguson_89a} and two outlying elliptical galaxies, NGC~1340 and IC~2006. 

Each galaxy was imaged in two {\it HST}/ACS filters, F475W and F850LP. The field of view (FOV) is $202\arcsec \times 202\arcsec$, with a pixel scale of $0\farcs049$. The two filters are roughly the same as the Sloan Digital Sky Survey (SDSS) $g$ and $z$ bands. Hereafter, they are referred as the $g$ or $z$ bands. For a full description of the ACSFCS and its data, we refer the reader to the survey description paper \citep{Jordan_07b} and the source catalog paper \citep{Jordan_15}. The basic properties of these 43 ETGs are listed in Table~\ref{galTab}. 
The galactic magnitudes are derived from the integration of the best-fit \sersic\ luminosity radial profiles, similar to the work of \citet{Ferrarese_06} for the ACSVCS samples. The stellar masses are from the estimates of $M_\star/L_z$ using the $g-z$ color and the models of \citet{BC03}, assuming a single age of 10~Gyr. These are the same galaxy quantities used in \citet{Turner_12}. 

% ------------ Tables ------------%
%\setcounter{table}{0}
%\begin{center}
\begin{deluxetable}{llccrr}
\tablewidth{0pt}
\tablecaption{ Global Properties of the Galaxies in ACSFCS 
 \label{galTab} }
\tablehead{
\colhead{Name} & 
\colhead{Other} & 
\colhead{$M_V$} & 
\colhead{$M_z$} & 
\colhead{$R_p$} & 
\colhead{$M_\star$}  
}
\decimalcolnumbers
\startdata
FCC~21  & NGC~1316 & $ -23.78 $ & $ -24.45 $ & $ 1.27 $ & $ 758.64 $ \\
FCC~213 & NGC~1399 & $ -22.77 $ & $ -23.49 $ & $ 0.00 $ & $ 263.02 $ \\
FCC~219 & NGC~1404 & $ -21.73 $ & $ -22.52 $ & $ 0.05 $ & $ 139.63 $ \\
NGC~1340 & ESO418-G005 & $ -21.59 $ & $ -22.22 $ & $ 1.69 $ & $  97.76 $ \\
FCC~167 & NGC~1380 & $ -21.83 $ & $ -22.46 $ & $ 0.21 $ & $  61.27 $ \\
FCC~276 & NGC~1427 & $ -20.99 $ & $ -21.68 $ & $ 0.27 $ & $  46.79 $ \\
FCC~147 & NGC~1374 & $ -20.79 $ & $ -21.47 $ & $ 0.24 $ & $  49.21 $ \\
IC~2006 & ESO~359-G007 & $ -20.00 $ & $ -20.75 $ & $ 1.15 $ & $  24.14 $ \\
FCC~83  & NGC~1351 & $ -20.50 $ & $ -21.18 $ & $ 0.59 $ & $  32.13 $ \\
FCC~184 & NGC~1387 & $ -20.95 $ & $ -21.75 $ & $ 0.11 $ & $  65.49 $ \\
FCC~63  & NGC~1339 & $ -19.98 $ & $ -20.75 $ & $ 1.32 $ & $  26.76 $ \\
FCC~193 & NGC~1389 & $ -20.26 $ & $ -20.90 $ & $ 0.13 $ & $  30.37 $ \\
FCC~153 & IC~1963 & $ -19.06 $ & $ -19.74 $ & $ 0.40 $ & $   8.68 $ \\
FCC~170 & NGC~1381 & $ -19.94 $ & $ -20.63 $ & $ 0.14 $ & $  19.70 $ \\
FCC~177 & NGC~1380A & $ -19.02 $ & $ -19.70 $ & $ 0.27 $ & $   6.67 $ \\
FCC~47  & NGC~1336 & $ -19.48 $ & $ -20.14 $ & $ 0.85 $ & $   9.33 $ \\
FCC~43  & IC~1919 & $ -18.94 $ & $ -19.51 $ & $ 1.25 $ & $   2.99 $ \\
FCC~190 & NGC~1380B & $ -18.92 $ & $ -19.62 $ & $ 0.13 $ & $   7.45 $ \\
FCC~310 & NGC~1460 & $ -19.10 $ & $ -19.75 $ & $ 0.70 $ & $   8.92 $ \\
FCC~148 & NGC~1375 & $ -19.35 $ & $ -19.92 $ & $ 0.23 $ & $  10.77 $ \\
FCC~249 & NGC~1419 & $ -19.13 $ & $ -19.77 $ & $ 0.73 $ & $   9.82 $ \\
FCC~255 & ESO358-G50 & $ -18.71 $ & $ -19.32 $ & $ 0.61 $ & $   3.59 $ \\
FCC~277 & NGC~1428 & $ -18.82 $ & $ -19.47 $ & $ 0.29 $ & $   7.55 $ \\
FCC~55  & ESO358-G06 & $ -18.34 $ & $ -18.98 $ & $ 0.85 $ & $   4.00 $ \\
FCC~152 & ESO358-G25 & $ -18.42 $ & $ -18.96 $ & $ 1.06 $ & $   3.49 $ \\
FCC~301 & ESO358-G59 & $ -18.46 $ & $ -19.08 $ & $ 0.50 $ & $   4.51 $ \\
FCC~335 & ESO359-G02 & $ -17.81 $ & $ -18.36 $ & $ 0.87 $ & $   1.27 $ \\
FCC~143 & NGC1373 & $ -18.50 $ & $ -19.16 $ & $ 0.26 $ & $   5.11 $ \\
FCC~95  & PGC~13084 & $ -17.90 $ & $ -18.52 $ & $ 0.50 $ & $   2.76 $ \\
FCC~136 & PGC~13230 & $ -17.56 $ & $ -18.15 $ & $ 0.28 $ & $   1.96 $ \\
FCC~182 & PGC~13343 & $ -17.71 $ & $ -18.37 $ & $ 0.11 $ & $   2.49 $ \\
FCC~204 & ESO358-G43 & $ -17.55 $ & $ -18.14 $ & $ 0.81 $ & $   1.80 $ \\
FCC~119 & PGC~13177 & $ -17.21 $ & $ -17.79 $ & $ 0.74 $ & $   1.06 $ \\
FCC~26  & ESO357-G25 & $ -17.32 $ & $ -17.54 $ & $ 1.06 $ & $   0.25 $ \\
FCC~90  & PGC~13058 & $ -17.44 $ & $ -17.81 $ & $ 0.59 $ & $   1.20 $ \\
FCC~106 & PGC~13146 & $ -17.38 $ & $ -17.96 $ & $ 0.58 $ & $   1.44 $ \\
FCC~19  & ESO301-G08 & $ -16.89 $ & $ -17.40 $ & $ 1.33 $ & $   0.34 $ \\
FCC~288 & ESO358-G56 & $ -16.92 $ & $ -17.47 $ & $ 0.63 $ & $   0.65 $ \\
FCC~202 & NGC~1396 & $ -17.12 $ & $ -17.68 $ & $ 0.02 $ & $   1.08 $ \\
FCC~324 & ESO358-G66 & $ -16.90 $ & $ -17.47 $ & $ 0.75 $ & $   0.57 $ \\
FCC~100 & PGC~13097 & $ -16.80 $ & $ -17.28 $ & $ 0.49 $ & $   0.86 $ \\
FCC~203 & ESO358-G42 & $ -16.97 $ & $ -17.52 $ & $ 0.32 $ & $   0.93 $ \\
FCC~303 & PGC~13758 & $ -17.34 $ & $ -17.88 $ & $ 0.70 $ & $   0.52 $ \\
\enddata
\tablecomments{ (1) FCC or primary designation (2) Alternate name 
(3) and (4) Absolute $V$ and $z$ magnitude; adopted from the unpublished measurements used in \citet{Turner_12}. (5) Projected distance from FCC~213 (NGC~1399); in Mpc. (6) Stellar mass ($10^9 M_\odot$), adopted from the measurements used in \citet{Turner_12}.}
\end{deluxetable}

%\end{center}
% ---------------------------------%

For the purpose of studying GCs, the images are sufficiently deep so that $\sim\!\!90\%$ of the GCs within the ACS FOV can be detected \citep{Cote_04}. We also used 16 blank high-latitude control fields with simulated Fornax galaxies introduced into the data to estimate the field-dependent contamination of background galaxies \citep[e.g.,][]{Peng_06}. 

\citet{Jordan_04} described the image analysis and point-source selection. \citet{Jordan_09} and \citet{Jordan_15} provided the GC catalogs with detection probabilities ($p_{GC}$) for the ACSVCS and ACSFCS data, respectively. The parameter $p_{GC}$, the probability of an object being a GC, is a function of an object's apparent magnitude ($m$), projected half-light radius ($r_h$), and the flux of the local background ($I_b$). Following the previous ACSVCS and ACSFCS papers, we select GCs by requiring $p_{GC} \ge 0.5$. 

In our \sn\ analysis, we will combine these data with the measurements from the ACSVCS sample \citep{P08}. The ACSVCS observed 100 ETGs in the Virgo Cluster with $B_T<16$, but the sample is only complete to $B_T<12.15$ ($M_B<-18.94$). It has an instrument setup identical to the ACSFCS, with data reduction and GC selection performed in the exact same way. \\

% ======================

\section{Counting Globular Clusters}
\label{count}

% ----------------------------------------
\subsection{Total Numbers of GCs}
\label{Ngc}

The stellar light of the galaxy creates a spatially varying completeness limit specific to each galaxy. Similar to the ACSVCS, most of the galaxies in the sample have lower mass, and thus low surface brightness and smaller spatial extent. For these galaxies, we find that a simple counting of GCs, corrected for contamination and the faint end of the GC luminosity function (GCLF), is sufficient. In each galaxy, we select GC candidates that are brighter than the 1$\sigma$ above the mean of its GCLF \citep[previously measured using the same data by][]{Villegas_10}. These are bright enough to be detected regardless of the underlying surface brightness of the galaxy.
To estimate the background contamination, we perform the same selection in control fields. Because the typical number of contaminants is less than 10 and the their spatial distributions are nearly uniform, we use the same background correction at all galactocentric distances. The average density and the 1$\sigma$ uncertainty of the contaminants are derived by the mean number and standard deviation of the selected candidates from 16 control fields. 

To count the total number of GCs, we order the selected GCs in a galaxy by their galactocentric distances. From the center to the outskirts, we add the number of $(1-n_B*dAi)/p_{GCi}$ at each (the {\it i}th) GC, where $n_B$, $dAi$, and $p_{GCi}$ are the mean density of contaminants, the area of the annulus between the ({\it i}-1)th and {\it i}th GC, and the $p_{GC}$ of this GC, respectively. 
For the low-mass ETGs ($M_z > -20$), the differential number densities at the outermost GCs are close to zero. Therefore, the total number of GCs in low-mass Fornax ETGs are just the cumulative numbers at the outermost GCs. 
For the massive ETGs ($M_z\leq-20$), we fit cumulative \sersic\ radial profiles and extrapolate them to infinity to derive the total numbers. For FCC~170 and FCC~193, because the results from profile fitting and simple sum are consistent with each other, we adopt the number from simple sum. 
The results of profile fitting and GC numbers are listed in Tables~\ref{tabSsc} and \ref{SNtab}. 

For the ACSVCS sample, we simply adopt the measurements published in \citet{P08}. In their work, the GC number of low-mass ETGs ($M_z > -19$) is counted by summing up the number of GCs, using a representative completeness limit for each galaxy, subtracting the contamination estimated by the control fields, and correcting for faint GCs using the GCLFs measured by \citet{Jordan_07a}, which is essentially the same technique as what we have done for the ACSFCS galaxies. For the massive Virgo ETGs, \citet{P08} fit the binned radial density profiles rather than the cumulative profiles, as we do here. 

We note that \citet{Harris_13} published GC numbers for most of our ACSFCS samples. Their numbers were adopted from \citet{Villegas_10} without corrections for either background contamination or completeness. 
Although the difference between most of our estimates and the \citet{Villegas_10} numbers is similar to the uncertainty, the numbers for low-mass ETGs presented in \citet{Harris_13} are systematically higher than ours. The mean difference is 10.21, which matches the expected contaminant numbers. Additionally, our more accurate measurements of the two most massive galaxies, FCC~213 and FCC~21, are significantly higher than that due to our accounting for GCs not in the ACS FOV.
\citet{Bassino_06} and \citet{Richtler_14} took wide field images of massive ETGs FCC~21, FCC~147, and FCC~184 and measured radial density profiles of their GC systems. Our measurements are consistent with theirs within the uncertainties. \\

% ----------------------------------------
\subsubsection{Special Cases}
\label{special}

Below, we describe some atypical or special cases: \\

\noindent\textit{FCC~213 (NGC~1399)}. 
A counterpart of M87 in Virgo. FCC~213 is the BCG of the Fornax Cluster. The FOV is insufficient to encompass its entire GC system, 
so we need to constrain the shape of its density profile using data at larger radii. 
Also observed as part of the ACSFCS, FCC~202 is only $4\farcm6$ away from FCC~213 . The GC candidates in the FCC~202 ACS field are evenly distributed throughout the image without any evidence of central concentration, implying that these GCs mostly belong to the GC system of the nearby massive galaxy. Therefore, we assume that the GCs outside of 6\re\ for FCC~202 ($\sim$half the size of the FOV) belong solely to the halo of FCC~213. 
We divide them into five bins of distance from GCC~213, with equal GC numbers in each bin, and use them as complementary density measurements outside the ACS image of FCC~213. 
In the even outer regions, \citet{Dirsch_03} conducted a wide-field study of NGC~1399 and sampled its GC system as far as $23\farcm$. We take the measurements between $4\farcm$ and $23\farcm$ galactocentric radius from the Table~3 in their paper and include them in our profile fitting. 
With 10 annuli of equal GC numbers, we fit the \sersic\ profile of its GC number density and integrate it to infinity to estimate the total number of FCC~213's GC system. 
\citet{Dirsch_03} found $6450\pm700$ GCs within a radius of $15\farcm$ (83~kpc), in agreement with ours in the same area. 

\noindent\textit{FCC~21 (NGC~1316)}. 
Unlike most local, massive ETGs, FCC~21 has experienced a relatively recent major merger that happened about 3~Gyr ago \citep{Goudfrooij_01b}. It contains a significant amount of intermediate-age GCs with relatively high metallicity, which possibly formed during the merger event \citep[e.g.][]{Goudfrooij_04}. This subpopulation of GCs is relatively faint and red. 
Most ETGs in our sample have GCLFs with a similar mean and standard deviation \citep{Villegas_10}.
However, the GCLF of FCC~21 is peaked at a significantly fainter magnitude with a much wider dispersion. Taking into account the brighter surface brightness of FCC~21, we only select out the GC candidates that are brighter than the mean of the GCLF to avoid problems with nondetection. Also, FCC~21 is in a special case in profile fitting. We find that the individual fitting of the blue and red GC populations (see Section~\ref{color} below) is better than the profile fitting of its entire GC system, and we adopt the total GC number to be the sum of these two populations. 
Furthermore, the closest companion of FCC~21, FCC~19, is a low-mass ETG with a small GC system. No excess of GCs is detected in its outskirts, indicating that it is not contaminated by FCC~21. 
We have checked the fitted GC number density profile of FCC~21 and find that it has fallen to zero at a radius smaller than the distance from FCC~19. 

\noindent\textit{FCC~219, FCC~184, FCC~182, and FCC~202}. The GC systems of these galaxies are contaminated by FCC~213. We subtract the contaminants from FCC~213 using its GC number density profile. 
For FCC~202, we only take into account the GCs inside 6\re, which corresponds to half the area of the FOV. 

\noindent\textit{FCC~148}. This is a low-mass ETG with a close massive companion, FCC~147. The projected distance between them is comparable to the size of the ACS FOV, and a quarter of its image is filled with the GCs from FCC~147. However, in the area opposite to the contaminated region, few GCs are detected. 
Therefore, when counting its GCs, we used only the central area dominated by the GC system of FCC~148 itself. We then divided the outer region equally into two sections, leaving most of the contaminants from FCC~147 in one of them. The GC number in the central area is counted after the subtraction of contaminants that are estimated from the fitted density profile. In the outer region, we measure the GC numbers in the half of the image with few contaminants, then double it. 
The total number of GCs is the sum of the counts from both the inner and outer areas. \\

% ----------------------------------------
\subsection{Separation of Blue and Red GCs}
\label{color}

Because the dividing line of the blue and red GC populations does not vary much among ETGs \citep{Peng_06}, we simply separate these two populations at $g-z=1.16$, which is the same cut used in \citet{P08}. 
We calculate the number in each GC subpopulation in the same way as for counting the total numbers but only extrapolate the density profiles for the ETGs for those galaxies with $M_z<-21$ (as well as IC~2006) when counting red GCs. 
If the profile fitting of the blue population has large uncertainty (due to being more spatially extended), we estimate the number by subtracting the number of the better-constrained red population from the total number. 

We note that the merger remnant FCC~21 contains a substantial fraction of intermediate-age GCs, especially in the outer regions. This means that the color distribution of its GC system has a significant red peak.There is still a bimodal color distribution, however \citep{Goudfrooij_01a,Goudfrooij_04}, and the color cut should probably be different from the one we use. \citet{Goudfrooij_04} suggested the blue and red peaks at $B-I$ = 1.5 and 1.8 and a possible separation at 1.65. However, because we do not want to add additional layers of interpretation with a movable color cut, we decided to apply a homogenous color cut on all galaxies and consider our measurements of blue and red GCs in FCC~21 as simply as a reference for comparing with the rest of the sample, keeping in mind the special nature of this galaxy.

Another special case is FCC~213. Part of the data used for its profile fitting are from \citet{Dirsch_03}. In their study, both the photometry systems and the color separation were different from ours. 
However, to constrain the profile shapes, we adopt the number densities of blue and red GCs at radii between $16\farcm$ to $23\farcm$ in \citet{Dirsch_03}. Accordingly, the results of the blue or red GCs of the galaxies that are contaminated by its GC systems are not on exactly the same scale as others. Fortunately, except for FCC~202, the contaminants from FCC~213 account for no more than 15\% of their GC numbers. Furthermore, the systematic errors caused by color calibrations should not affect our results qualitatively. \\

% =======================

\section{Results}
\label{result}

% ------------ Tables ------------%

\begin{deluxetable*}{lccccccccc}
\tabletypesize{\tiny}
\tablewidth{0pt}
\tablecaption{ The Cumulative \sersic\ Profile Fitting Results of the GC Systems in ACSFCS Galaxies
 \label{tabSsc} }
\tablehead{
\colhead{ID} & 
\colhead{$N_{e,tot}$} & 
\colhead{$R_{e,tot}$ (arcsec)} & 
\colhead{$n_{tot}$} &
\colhead{$N_{e,blue}$} &
\colhead{$R_{e,blue}$ (arcsec)} &
\colhead{$n_{blue}$} &
\colhead{$N_{e,red}$} &
\colhead{$R_{e,red}$ (arcsec)} &
\colhead{$n_{red}$}  
}
\decimalcolnumbers
\startdata
F21   & $ 0.0343 \pm 0.0007 $ & $ 77.87  \pm 1.99   $ & $ 0.40 \pm 0.01 $ & $ 0.0189 \pm 0.0013 $ & $ 93.82 \pm 15.10 $ & $ 0.41 \pm 0.04 $ & $ 0.0086 \pm 0.0022 $ & $ 107.49 \pm 27.10 $ & $ 0.82 \pm 0.18 $ \\  
F213  & $ 0.0010 \pm 0.0018 $ & $ 655.78 \pm 233.43 $ & $ 3.34 \pm 0.47 $ & $ 0.0004 \pm 0.0004 $ & $ 730.56 \pm 277.39 $ & $ 2.88 \pm 0.67 $ & $ 0.0011 \pm 0.0012 $ & $ 551.21 \pm 163.73 $ & $ 2.59 \pm 0.31 $ \\  
N1340 & $ 0.0082 \pm 0.0009 $ & $ 56.91  \pm 3.91   $ & $ 1.21 \pm 0.10 $ & $ 0.0037 \pm 0.0008 $ & $ 83.73 \pm 14.26 $ & $ 1.63 \pm 0.22 $ & $ 0.0036 \pm 0.0009 $ & $ 37.88 \pm 6.47 $ & $ 0.85 \pm 0.31 $ \\
F219  & $ 0.0160 \pm 0.0008 $ & $ 42.45  \pm 1.18   $ & $ 0.76 \pm 0.05 $ & $ 0.0065 \pm 0.0011 $ & $ 46.62 \pm 15.27 $ & $ 0.64 \pm 0.24 $ & $ 0.0099 \pm 0.0010 $ & $ 39.24 \pm 2.12 $ & $ 0.85 \pm 0.10 $ \\  
F167  & $ 0.0149 \pm 0.0019 $ & $ 53.45  \pm 9.74   $ & $ 0.97 \pm 0.16 $ & $ 0.0058 \pm 0.0007 $ & $ 44.70 \pm 3.38 $ & $ 0.83 \pm 0.12 $ & $ 0.0081 \pm 0.0014 $ & $ 63.18 \pm 15.19 $ & $ 1.15 \pm 0.20 $ \\  
F276  & $ 0.0117 \pm 0.0014 $ & $ 52.25  \pm 6.23   $ & $ 1.35 \pm 0.15 $ & $ 0.0044 \pm 0.0007 $ & $ 74.94 \pm 9.56 $ & $ 1.47 \pm 0.15 $ & $ 0.0079 \pm 0.0012 $ & $ 37.14 \pm 3.20 $ & $ 1.26 \pm 0.16 $ \\  
F147  & $ 0.0073 \pm 0.0011 $ & $ 67.48  \pm 7.03   $ & $ 1.48 \pm 0.14 $ & $ 0.0050 \pm 0.0008 $ & $ 58.40 \pm 6.67 $ & $ 0.99 \pm 0.14 $ & $ -- $ & $ -- $ & $ -- $  \\  
I2006  & $ 0.0111 \pm 0.0013 $ & $ 30.58  \pm 1.89   $ & $ 0.93 \pm 0.13 $ & $ 0.0023 \pm 0.0012 $ & $ 54.08 \pm 32.06 $ & $ 2.18 \pm 0.77 $ & $ 0.0079 \pm 0.0012 $ & $ 29.14 \pm 2.74 $ & $ 0.61 \pm 0.16 $ \\
F83   & $ 0.0137 \pm 0.0018 $ & $ 36.93  \pm 2.46   $ & $ 1.88 \pm 0.14 $ & $ 0.0048 \pm 0.0014 $ & $ 51.68 \pm 9.53 $ & $ 2.55 \pm 0.36 $ & $ 0.0077 \pm 0.0022 $ & $ 33.21 \pm 11.81 $ & $ 1.59 \pm 0.48 $ \\  
F184  & $ 0.0131 \pm 0.0006 $ & $ 51.39  \pm 4.29   $ & $ 0.40 \pm 0.04 $ & $ 0.0044 \pm 0.0004 $ & $ 43.60 \pm 5.84 $ & $ 0.47 \pm 0.10 $ & $ 0.0085 \pm 0.0004 $ & $ 56.22 \pm 6.91 $ & $ 0.40 \pm 0.05 $ \\  
F63   & $ 0.0177 \pm 0.0012 $ & $ 32.73  \pm 1.15   $ & $ 1.03 \pm 0.07 $ & $ 0.0084 \pm 0.0015 $ & $ 41.07 \pm 11.77 $ & $ 0.95 \pm 0.28 $ & $ 0.0112 \pm 0.0017 $ & $ 24.60 \pm 2.02 $ & $ 1.11 \pm 0.19 $ \\  
F47   & $ 0.0118 \pm 0.0014 $ & $ 40.57  \pm 2.59   $ & $ 1.58 \pm 0.13 $ & $ -- $ & $ -- $ & $ -- $ & $ 0.0168 \pm 0.0027 $ & $ 20.25 \pm 2.52 $ & $ 0.89 \pm 0.23 $ \\  
\enddata
\tablecomments{ (1) The ID of galaxies. 
(2) (5) (8) The normalization factor of the \sersic\ profiles at 1\re; for total, blue, and red GCs respectively. 
(3) (6) (9) The fitted \re\ of total, blue, and red GCs. 
(4) (7) (10) The \sersic\ index of total, blue, and red GCs. }
\end{deluxetable*}

% ---------------------------------%
%\setcounter{table}{1}
%\begin{center}
%\tabletypesize{\tiny}
\begin{deluxetable*}{lccccccr}
\tabletypesize{\scriptsize}
\tablewidth{0pt}
\tablecaption{ The Numbers and Formation Efficiencies of GCs in ACSFCS Galaxies
 \label{SNtab} }
\tablehead{
\colhead{ID} & 
\colhead{$N_{GC}$} & 
\colhead{$N_{GC,blue}$} & 
\colhead{$N_{GC,red}$} & 
\colhead{$S_N$} & 
\colhead{$S_{N,z}$} & 
\colhead{$T$} &
\colhead{$f_{red}$} 
}
\decimalcolnumbers
\startdata
FCC~21  & $ 2541.66 \pm 514.16  $ & $1442.2 \pm 406.5 $ & $ 1099.49 \pm 314.83$ & $ 0.54 \pm 0.02 $ & $ 0.29 \pm 0.01 $ & $  2.35 \pm  1.03 $ & $ 0.19 \pm 0.23 $ \\
FCC~213 & $ 9693.0 \pm 3263.3$ & $4234.8 \pm 813.7 $ & $6655.4 \pm 1841.3$ & $ 7.49 \pm 2.52 $ & $ 3.87 \pm 1.30 $ & $ 36.85 \pm 23.59 $ & $ 0.61 \pm 0.20 $ \\
FCC~219 & $  310.6 \pm  8.16 $ & $ 141.7 \pm  37.2 $ & $ 171.7 \pm   8.3 $ & $ 0.62 \pm 0.01 $ & $ 0.30 \pm 0.00 $ & $  2.22 \pm  0.88 $ & $ 0.54 \pm 0.07 $ \\
FCC~167 & $  503.9 \pm  76.6 $ & $ 129.5 \pm   9.0 $ & $ 374.3 \pm  77.2 $ & $ 0.92 \pm 0.14 $ & $ 0.52 \pm 0.07 $ & $  8.22 \pm  1.25 $ & $ 0.74 \pm 0.19 $ \\
FCC~276 & $  440.4 \pm  39.7 $ & $ 295.5 \pm  40.9 $ & $ 145.0 \pm  10.0 $ & $ 1.76 \pm 0.15 $ & $ 0.93 \pm 0.08 $ & $  9.41 \pm  5.20 $ & $ 0.32 \pm 0.03 $ \\
FCC~147 & $  473.7 \pm  41.8 $ & $ 206.8 \pm  22.6 $ & $ 266.9 \pm  47.5 $ & $ 2.26 \pm 0.20 $ & $ 1.22 \pm 0.10 $ & $  9.62 \pm  4.12 $ & $ 0.56 \pm 0.11 $ \\
FCC~83  & $  298.1 \pm  13.6 $ & $ 172.4 \pm  26.4 $ & $ 125.7 \pm  22.6 $ & $ 1.87 \pm 0.08 $ & $ 1.00 \pm 0.04 $ & $  9.27 \pm  4.71 $ & $ 0.42 \pm 0.07 $ \\
FCC~184 & $  299.0 \pm  30.9 $ & $  75.6 \pm  10.3 $ & $ 233.8 \pm  39.3 $ & $ 1.23 \pm 0.12 $ & $ 0.59 \pm 0.06 $ & $  4.56 \pm  2.02 $ & $ 0.75 \pm 0.16 $ \\
FCC~63  & $  231.8 \pm   6.9 $ & $ 167.3 \pm  32.2 $ & $  81.0 \pm  10.0 $ & $ 2.34 \pm 0.06 $ & $ 1.15 \pm 0.03 $ & $  8.66 \pm  3.56 $ & $ 0.32 \pm 0.05 $ \\
FCC~193 & $   38.6 \pm   7.1 $ & $  21.4 \pm   5.4 $ & $  12.6 \pm   4.8 $ & $ 0.30 \pm 0.05 $ & $ 0.16 \pm 0.03 $ & $  1.27 \pm  0.57 $ & $ 0.37 \pm 0.16 $ \\
FCC~170 & $   60.0 \pm   8.9 $ & $  52.6 \pm   8.2 $ & $   7.8 \pm   3.8 $ & $ 0.63 \pm 0.09 $ & $ 0.33 \pm 0.04 $ & $  3.04 \pm  1.70 $ & $ 0.12 \pm 0.06 $ \\
FCC~153 & $   48.1 \pm   8.0 $ & $  45.2 \pm   7.6 $ & $   3.9 \pm   2.5 $ & $ 1.13 \pm 0.18 $ & $ 0.60 \pm 0.10 $ & $  5.53 \pm  2.87 $ & $ 0.07 \pm 0.05 $ \\
FCC~177 & $   66.6 \pm   9.2 $ & $  61.7 \pm   8.8 $ & $   4.8 \pm   2.8 $ & $ 1.63 \pm 0.22 $ & $ 0.87 \pm 0.12 $ & $  9.98 \pm  6.30 $ & $ 0.07 \pm 0.04 $ \\
FCC~47  & $  286.6 \pm  13.4 $ & $ 208.2 \pm  16.6 $ & $  78.5 \pm   9.8 $ & $ 4.61 \pm 0.21 $ & $ 2.51 \pm 0.11 $ & $ 30.71 \pm 20.74 $ & $ 0.27 \pm 0.03 $ \\
FCC~43  & $   31.3 \pm   6.6 $ & $  27.2 \pm   6.0 $ & $   4.4 \pm   3.0 $ & $ 0.83 \pm 0.17 $ & $ 0.49 \pm 0.10 $ & $ 10.46 \pm 12.93 $ & $ 0.14 \pm 0.10 $ \\
FCC~190 & $  141.2 \pm  13.2 $ & $ 121.6 \pm  12.1 $ & $  14.1 \pm   5.3 $ & $ 3.79 \pm 0.35 $ & $ 2.00 \pm 0.18 $ & $ 18.94 \pm 10.30 $ & $ 0.10 \pm 0.04 $ \\
FCC~310 & $   28.6 \pm   6.7 $ & $  25.4 \pm   5.9 $ & $   3.4 \pm   3.3 $ & $ 0.65 \pm 0.15 $ & $ 0.35 \pm 0.08 $ & $  3.20 \pm  1.82 $ & $ 0.11 \pm 0.11 $ \\
FCC~249 & $  146.2 \pm  13.4 $ & $ 123.0 \pm  12.3 $ & $  22.9 \pm   5.6 $ & $ 3.23 \pm 0.29 $ & $ 1.79 \pm 0.16 $ & $ 14.87 \pm  7.11 $ & $ 0.15 \pm 0.04 $ \\
FCC~148 & $   22.6 \pm  10.8 $ & $  22.5 \pm   8.2 $ & $   0.2 \pm  13.6 $ & $ 0.41 \pm 0.19 $ & $ 0.24 \pm 0.11 $ & $  2.10 \pm  1.42 $ & $ 0.00 \pm 0.59 $ \\
FCC~255 & $   68.8 \pm   9.2 $ & $  55.5 \pm   8.3 $ & $  12.9 \pm   4.3 $ & $ 2.25 \pm 0.30 $ & $ 1.28 \pm 0.17 $ & $ 19.13 \pm 17.27 $ & $ 0.18 \pm 0.06 $ \\
FCC~277 & $   36.9 \pm   6.9 $ & $  28.3 \pm   6.0 $ & $   9.7 \pm   3.4 $ & $ 1.08 \pm 0.20 $ & $ 0.59 \pm 0.11 $ & $  4.88 \pm  2.38 $ & $ 0.25 \pm 0.10 $ \\
FCC~55  & $   29.2 \pm   6.1 $ & $  21.6 \pm   5.2 $ & $   7.9 \pm   3.2 $ & $ 1.34 \pm 0.28 $ & $ 0.74 \pm 0.15 $ & $  7.30 \pm  4.33 $ & $ 0.26 \pm 0.12 $ \\
FCC~152 & $    9.3 \pm   3.6 $ & $   8.5 \pm   3.4 $ & $   1.2 \pm   1.2 $ & $ 0.39 \pm 0.15 $ & $ 0.24 \pm 0.09 $ & $  2.66 \pm  1.91 $ & $ 0.12 \pm 0.13 $ \\
FCC~301 & $   15.0 \pm   5.3 $ & $  13.8 \pm   4.7 $ & $   1.9 \pm   2.8 $ & $ 0.61 \pm 0.21 $ & $ 0.34 \pm 0.12 $ & $  3.33 \pm  2.10 $ & $ 0.11 \pm 0.18 $ \\
FCC~335 & $    9.9 \pm   3.6 $ & $   9.9 \pm   3.6 $ & $   0.0 \pm   0.0 $ & $ 0.74 \pm 0.26 $ & $ 0.45 \pm 0.16 $ & $  7.82 \pm  7.79 $ & $ 0.00 \pm 0.00 $ \\
NGC~134 & $  347.5 \pm  20.8 $ & $ 288.1 \pm  22.2 $ & $  59.4 \pm   7.6 $ & $ 0.79 \pm 0.04 $ & $ 0.44 \pm 0.02 $ & $  3.55 \pm  1.64 $ & $ 0.17 \pm 0.02 $ \\
FCC~143 & $   49.7 \pm   8.1 $ & $  44.7 \pm   7.6 $ & $   5.3 \pm   2.7 $ & $ 1.96 \pm 0.32 $ & $ 1.06 \pm 0.17 $ & $  9.71 \pm  5.34 $ & $ 0.10 \pm 0.05 $ \\
FCC~95  & $    9.8 \pm   3.9 $ & $   9.7 \pm   3.6 $ & $  -0.1 \pm   1.7 $ & $ 0.68 \pm 0.26 $ & $ 0.38 \pm 0.15 $ & $  3.57 \pm  2.29 $ & $-0.01 \pm 0.17 $ \\
FCC~136 & $   16.4 \pm   4.6 $ & $  15.6 \pm   4.5 $ & $   1.1 \pm   1.2 $ & $ 1.54 \pm 0.43 $ & $ 0.89 \pm 0.25 $ & $  8.32 \pm  4.89 $ & $ 0.06 \pm 0.07 $ \\
FCC~182 & $   37.5 \pm   8.3 $ & $  38.7 \pm   7.6 $ & $   1.5 \pm   3.0 $ & $ 3.06 \pm 0.67 $ & $ 1.67 \pm 0.36 $ & $ 15.06 \pm  8.12 $ & $ 0.03 \pm 0.07 $ \\
FCC~204 & $   12.0 \pm   4.2 $ & $  11.6 \pm   4.0 $ & $   0.5 \pm   1.2 $ & $ 1.14 \pm 0.39 $ & $ 0.65 \pm 0.22 $ & $  6.63 \pm  4.53 $ & $ 0.04 \pm 0.10 $ \\
FCC~119 & $   11.6 \pm   4.6 $ & $  10.0 \pm   4.1 $ & $   1.7 \pm   2.2 $ & $ 1.50 \pm 0.59 $ & $ 0.88 \pm 0.34 $ & $ 10.88 \pm  8.75 $ & $ 0.14 \pm 0.19 $ \\
FCC~90  & $   16.0 \pm   4.7 $ & $  12.7 \pm   4.2 $ & $   3.6 \pm   2.1 $ & $ 1.68 \pm 0.49 $ & $ 1.19 \pm 0.35 $ & $ 13.29 \pm  8.98 $ & $ 0.21 \pm 0.14 $ \\
FCC~26  & $   14.8 \pm   4.3 $ & $  14.9 \pm   4.3 $ & $   0.0 \pm   0.0 $ & $ 1.74 \pm 0.50 $ & $ 1.42 \pm 0.41 $ & $ 58.75 \pm 138.32$ & $ 0.00 \pm 0.00 $ \\
FCC~106 & $   10.5 \pm   4.5 $ & $   8.8 \pm   4.0 $ & $   2.9 \pm   2.1 $ & $ 1.17 \pm 0.49 $ & $ 0.68 \pm 0.29 $ & $  7.25 \pm  5.57 $ & $ 0.24 \pm 0.20 $ \\
FCC~19  & $    7.6 \pm   3.6 $ & $   8.4 \pm   3.5 $ & $   0.0 \pm   0.0 $ & $ 1.32 \pm 0.62 $ & $ 0.82 \pm 0.38 $ & $ 22.10 \pm 36.02 $ & $ 0.00 \pm 0.00 $ \\
FCC~202 & $    7.2 \pm   8.7 $ & $   4.7 \pm   6.0 $ & $   2.5 \pm  10.6 $ & $ 1.02 \pm 1.23 $ & $ 0.60 \pm 0.73 $ & $  6.66 \pm  9.00 $ & $ 0.34 \pm 1.52 $ \\
FCC~324 & $   12.9 \pm   4.4 $ & $  13.1 \pm   4.4 $ & $   0.0 \pm   0.0 $ & $ 2.25 \pm 0.76 $ & $ 1.32 \pm 0.45 $ & $ 22.61 \pm 20.11 $ & $ 0.00 \pm 0.00 $ \\
FCC~288 & $   13.1 \pm   5.0 $ & $  11.1 \pm   4.4 $ & $   2.1 \pm   2.6 $ & $ 2.22 \pm 0.84 $ & $ 1.33 \pm 0.51 $ & $ 19.83 \pm 17.92 $ & $ 0.16 \pm 0.20 $ \\
FCC~303 & $   12.8 \pm   4.2 $ & $  13.3 \pm   4.2 $ & $   0.0 \pm   0.0 $ & $ 1.48 \pm 0.48 $ & $ 0.89 \pm 0.29 $ & $ 24.30 \pm 38.50 $ & $ 0.00 \pm 0.00 $ \\
FCC~203 & $   21.6 \pm   5.7 $ & $  19.9 \pm   5.3 $ & $   1.8 \pm   2.3 $ & $ 3.50 \pm 0.92 $ & $ 2.10 \pm 0.55 $ & $ 23.22 \pm 15.13 $ & $ 0.08 \pm 0.10 $ \\
FCC~100 & $   23.6 \pm   5.7 $ & $  23.0 \pm   5.6 $ & $   0.2 \pm   1.3 $ & $ 4.49 \pm 1.07 $ & $ 2.87 \pm 0.69 $ & $ 27.26 \pm 14.51 $ & $ 0.01 \pm 0.05 $ \\
IC~2006 & $  121.7 \pm   6.4 $ & $  55.0 \pm   8.9 $ & $  66.7 \pm   6.2 $ & $ 1.21 \pm 0.06 $ & $ 0.60 \pm 0.03 $ & $  5.04 \pm  2.43 $ & $ 0.54 \pm 0.05 $ \\
\enddata
\tablecomments{ (1) The ID of galaxies; (2) Total number of globular clusters; (3) Total number of blue globular clusters; (4) Total number of red globular clusters; (5) Specific frequency; (6) Specific frequency in $z$ bandpass; (7) $N_{GC}$ normalized to stellar mass of $10^9 M_\odot$; (8) Fraction of red GCs. }
\end{deluxetable*}

%\end{center}
% ---------------------------------%

\begin{deluxetable}{lccccc}
\tablewidth{0pt}
\tablecaption{ \snz\ within 1\re\ of Massive Galaxies in ACSFCS and ACSVCS. 
 \label{tabInOut} }
\tablehead{
\colhead{ID} & 
\colhead{R$_{e,gal}$ ($"$)} & 
\colhead{n$_{gal}$} & 
\colhead{$M_{z,In}$} & 
\colhead{$N_{GC,In}$} & 
\colhead{$S_{N,z,In}$}  
}
\decimalcolnumbers
\startdata
F21   & $ 146.03 $ & $ 6.29 $ & $ -23.69 $ & $ 950.76 \pm 192.33 $ & $ 0.31 \pm 0.06$  \\ 
F213  & $ 114.58 $ & $ 7.51 $ & $ -22.73 $ & $ 1083.65\pm 364.82 $ & $ 0.87 \pm 0.29$  \\ 
F219  & $ 22.91  $ & $ 3.85 $ & $ -21.76 $ & $ 65.11  \pm 8.91   $ & $ 0.12 \pm 0.01$  \\ 
F167  & $ 30.33  $ & $ 4.0  $ & $ -21.70 $ & $ 125.77 \pm 12.34  $ & $ 0.26 \pm 0.02$  \\ 
F276  & $ 35.70  $ & $ 5.39 $ & $ -20.92 $ & $ 146.23 \pm 13.36  $ & $ 0.62 \pm 0.05$  \\ 
F147  & $ 29.20  $ & $ 5.13 $ & $ -20.71 $ & $ 97.48  \pm 10.89  $ & $ 0.50 \pm 0.05$  \\ 
F83   & $ 26.56  $ & $ 4.98 $ & $ -20.42 $ & $ 106.09 \pm 11.34  $ & $ 0.71 \pm 0.07$  \\ 
F184  & $ 21.37  $ & $ 4.14 $ & $ -20.99 $ & $ 17.47  \pm 4.79   $ & $ 0.06 \pm 0.01$  \\ 
F63   & $ 14.67  $ & $ 3.0  $ & $ -19.99 $ & $ 39.06  \pm 6.92   $ & $ 0.39 \pm 0.06$  \\ 
F193  & $ 11.97  $ & $ 2.72 $ & $ -20.14 $ & $ 9.63   \pm 3.38   $ & $ 0.08 \pm 0.02$  \\ 
F170  & $ 6.83   $ & $ 2.0  $ & $ -19.87 $ & $ -1.19  \pm 0.0    $ & $ -0.01\pm 0.0 $  \\ 
F47   & $ 27.65  $ & $ 3.08 $ & $ -19.38 $ & $ 103.04 \pm 11.19  $ & $ 1.81 \pm 0.19$  \\ 
N1340 & $ 33.21  $ & $ 3.78 $ & $ -21.46 $ & $ 92.37  \pm 10.71  $ & $ 0.23 \pm 0.02$  \\ 
I2006 & $ 18.94  $ & $ 3.50 $ & $ -19.99 $ & $ 26.94  \pm 5.87   $ & $ 0.27 \pm 0.05$  \\ 
V1978 & $ 98.48  $ & $ 4.85 $ & $ -22.66 $ & $ 726.75 \pm 29.93  $ & $ 0.62 \pm 0.02$  \\
V1632 & $ 84.37  $ & $ 7.60 $ & $ -21.57 $ & $ 376.59 \pm 21.76  $ & $ 0.88 \pm 0.05$  \\
V1231 & $ 18.64  $ & $ 3.22 $ & $ -20.82 $ & $ 38.13  \pm 6.91   $ & $ 0.17 \pm 0.03$  \\ 
V2095 & $ 14.29  $ & $ 3.96 $ & $ -20.19 $ & $ 13.20  \pm 3.96   $ & $ 0.11 \pm 0.03$  \\ 
V1154 & $ 29.50  $ & $ 5.07 $ & $ -21.03 $ & $ 47.95  \pm 7.74   $ & $ 0.18 \pm 0.02$  \\ 
V1062 & $ 14.21  $ & $ 3.14 $ & $ -20.58 $ & $ 15.49  \pm 4.29   $ & $ 0.09 \pm 0.02$  \\ 
V2092 & $ 31.45  $ & $ 5.09 $ & $ -20.88 $ & $ 22.27  \pm 5.35   $ & $ 0.09 \pm 0.02$  \\ 
V369  & $ 7.74   $ & $ 2.76 $ & $ -19.65 $ & $ 2.51   \pm 1.73   $ & $ 0.03 \pm 0.02$  \\ 
V759  & $ 27.27  $ & $ 3.53 $ & $ -20.65 $ & $ 59.43  \pm 8.54   $ & $ 0.32 \pm 0.04$  \\ 
V1692 & $ 9.34   $ & $ 2.73 $ & $ -20.28 $ & $ 5.95   \pm 2.91   $ & $ 0.04 \pm 0.02$  \\ 
V2000 & $ 9.44   $ & $ 4.45 $ & $ -19.90 $ & $ 18.06  \pm 4.78   $ & $ 0.19 \pm 0.05$  \\ 
V685  & $ 11.20  $ & $ 3.21 $ & $ -20.28 $ & $ 11.92  \pm 3.95   $ & $ 0.09 \pm 0.03$  \\ 
V1664 & $ 15.91  $ & $ 4.33 $ & $ -20.21 $ & $ 29.26  \pm 6.10   $ & $ 0.23 \pm 0.05$  \\ 
V654  & $ 20.78  $ & $ 4.03 $ & $ -19.83 $ & $ 7.20   \pm 2.92   $ & $ 0.08 \pm 0.03$  \\ 
V944  & $ 11.61  $ & $ 3.35 $ & $ -19.90 $ & $ 8.35   \pm 3.37   $ & $ 0.09 \pm 0.03$  \\ 
V1938 & $ 13.36  $ & $ 4.11 $ & $ -20.10 $ & $ 7.20   \pm 2.92   $ & $ 0.06 \pm 0.02$  \\ 
V1279 & $ 12.01  $ & $ 2.15 $ & $ -19.96 $ & $ 13.10  \pm 4.12   $ & $ 0.13 \pm 0.04$  \\ 
V1720 & $ 27.22  $ & $ 4.46 $ & $ -19.92 $ & $ 13.10  \pm 4.12   $ & $ 0.14 \pm 0.04$  \\ 
V355  & $ 8.83   $ & $ 3.69 $ & $ -19.65 $ & $ -1.19  \pm 0.0    $ & $ -0.01\pm 0.0 $  \\ 
V1619 & $ 9.98   $ & $ 1.66 $ & $ -19.43 $ & $ 7.72   \pm 3.25   $ & $ 0.12 \pm 0.05$  \\ 
V1883 & $ 25.02  $ & $ 4.57 $ & $ -19.97 $ & $ 27.82  \pm 5.87   $ & $ 0.28 \pm 0.05$  \\ 
V1242 & $ 16.67  $ & $ 3.36 $ & $ -19.53 $ & $ 26.23  \pm 5.71   $ & $ 0.40 \pm 0.08$  \\ 
V784  & $ 13.67  $ & $ 3.00 $ & $ -19.50 $ & $ 10.18  \pm 3.82   $ & $ 0.16 \pm 0.06$  \\ 
V778  & $ 5.19   $ & $ 3.03 $ & $ -19.27 $ & $ -1.19  \pm 0.0    $ & $ -0.02\pm 0.0 $  \\ 
V828  & $ 10.50  $ & $ 2.20 $ & $ -19.31 $ & $ 1.19   \pm 1.68   $ & $ 0.02 \pm 0.03$  \\ 
V1250 & $ 11.78  $ & $ 7.97 $ & $ -19.34 $ & $ 4.76   \pm 2.38   $ & $ 0.08 \pm 0.04$  \\ 
V1630 & $ 13.11  $ & $ 2.24 $ & $ -19.30 $ & $ 16.90  \pm 4.64   $ & $ 0.32 \pm 0.08$  \\ 
V1025 & $ 10.12  $ & $ 3.71 $ & $ -19.81 $ & $ 10.90  \pm 3.79   $ & $ 0.12 \pm 0.04$  \\ 
\enddata
\tablecomments{(1) ID of massive galaxies. (2) and (3) Galactic \re\ in the unit of arcseconds and \sersic index; adopted from measurements used in \citet{Turner_12} and \citet{Ferrarese_06} for ACSFCS and ACSVCS galaxies, respectively. 
(4) z-band absolute magnitude within 1\re\ of galactic main bodies. 
(5) and (6) The total number of GCs and \snz\ within 1\re\ of the stars, respectively. }

\end{deluxetable}

% ---------------------------------%

Traditionally, \sn\ is defined as the number of GCs per unit galactic stellar $V$-band luminosity 
 \citep{Harris_vdB_81}. 
\citet{P08} introduced the modified parameter, \snz, (Equation~(\ref{S_Nz})) that is similar to \sn, but normalized to the galaxy luminosity of $M_z =-15$, with the $z$-band being a redder bandpass that more accurately reflects the stellar mass.
Similarly, \citet{Zepf_Ashman_93} defined $T$ as the number of GCs per $10^9~M_\odot$ of galactic stellar mass (Equation~(\ref{T})), which has the advantage of comparing galaxies with different stellar mass-to-light ratios ($M_\star/L$): 

%% -------- Eq.s --------- %
%\begin{equation}
%S_N = N_{GC} \times 10^{ 0.4(M_V+15) } \\ 
%\label{S_N}
%\end{equation}
% -------- Eq.s --------- %
\begin{equation}
S_{N,z} = N_{GC} \times 10^{ 0.4(M_z+15) } \\ 
\label{S_Nz}
\end{equation}
% -------- Eq.s --------- %
\begin{equation}
T = N_{GC} / ( M_{G^\star} / 10^9~M_\odot ) \\ 
\label{T}
\end{equation}
% -------------------------- %

All of these parameters indicate the GC formation (and survival) efficiency relative to that of field stars, and they are listed in Table~\ref{SNtab}. Because $z$-band magnitude is a better indicator of stellar mass than $V$-band, and we have actually measured $M_z$ for our sample galaxies, \snz\ is better than \sn\ for our purposes. Furthermore, our samples are all ETGs, which have similar $M_\star/L$. We have tested that in all the analysis in this work, \snz\ and $T$ have similar trends in the relation to other parameters. Because the measurement of stellar mass relies on stellar population models that would bring in additional systematic uncertainties, we primarily use \snz\ in our analysis.

% ---------------------------------------- 
\subsection{Dependence on Galactic Mass}
\label{R_mass}

\begin{figure}
\epsscale{1.3}
\plotone{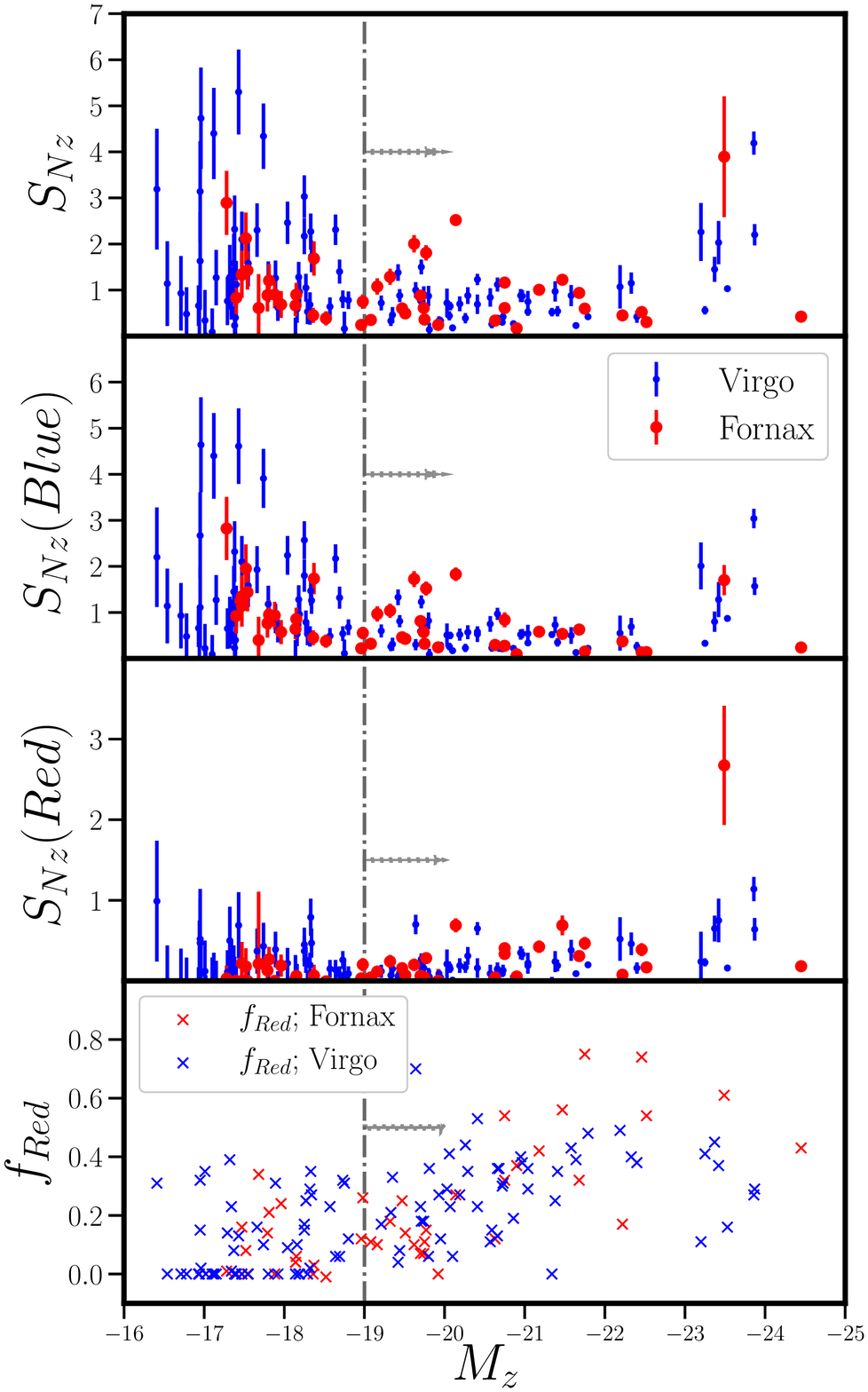}
\caption{ Relation between \snz\ and $z$-band absolute magnitudes of the 143 ETGs from ACSFCS (red circles) and ACSVCS (blue dots) samples. The \snz\ in the first three panels are measured by the total GC population, blue GCs, and red GCs in galaxies, respectively. 
The \snz\ of the entire GC and blue GC populations have similar trends with \Mz. They have a roughly constant value for intermediate-mass ETGs and increases with galactic luminosity at the bright end ($M_z<-23$). They also systematically rise at faint end, but with large scatter. 
The vertical dotted-dashed lines show the dividing line for low-mass ETGs defined in \citet{P08}, but we redefine it as $M_z<-20$ (illustrated by the arrows) in this work. 
The \snz\ values of red GCs are nearly all below $\sim$1, slightly increasing with \Mz\ among intermediate-mass ETGs. 
In the bottom panel, the red and blue crosses illustrate the relation between \fRed\ (y-axis on the right) and \Mz\ in the Fornax and Virgo Clusters.  
The low-mass ETGs fainter than $M_z=-20$ have relatively low \fRed\ below 0.4, and some of the faintest ETGs have no red GC. Here \fRed\ is positively correlated with \Mz\ in the intermediate-mass range and flattens or decreases with \Mz\ at the high-mass end. 
 \label{SN_M} }
\end{figure}

Figure~\ref{SN_M} displays the relation between \snz\ and \Mz\ of the 143 ETGs in the ACSFCS and ACSVCS. The measurements for Virgo galaxies are adopted directly from \citet{P08}. Red circles and blue dots represent the ETGs from Fornax and Virgo, respectively. From top to bottom, the y-axis is \snz\ measured for the entire GC system, the blue GC population, and the red GC population, respectively. 

% ------------- PLOTs --------------- %
% --------------------------------------- %

The distributions of all three parameters are similar in Fornax and Virgo. 
Specifically, the \snz\ for all GCs is roughly constant with some scatter, but largely below \snz$=1.5$ among intermediate-mass ETGs, and it increases with galactic luminosity at the bright end ($M_z<-23$). It also systematically rises at the faint end, but with large scatter. While the lowest values reach zero, the highest values are comparable or even higher than the highest \snz\ at the massive end in their host clusters. 

However, the U shape of this relation is mainly driven by the ETGs in Virgo. This may simply be a result of the smaller sample size in Fornax. At the bright end, only two ETGs from Fornax have \Mz\ brighter than $-23$. While the BCG FCC~213 has a high \snz\ that follows the U-shaped distribution in Virgo, the other one, FCC~21, is a post-starburst galaxy and has a low \snz\ close to zero. This low \snz\ is possibly due to the high luminosity produced by its relatively young stellar population. 

At the faint end, because Fornax is 3.5~Mpc farther than Virgo, the ACSFCS did not sample the ETGs fainter than $M_z\sim-17$. 
In the range of magnitudes that overlap, the ETGs with the highest \snz\ at each magnitude are mostly from Virgo. In Virgo, there are galaxies with \snz\ higher than that for the BCG (M87), while this is not the case in Fornax. 

\citet{P08} drew the dividing line between low- and intermediate-mass galaxies at \Mz=$-19$~mag (the vertical dotted-dashed lines in Figure~1) when only using the ETGs from Virgo. 
However, the scatter starts to increase upward at $M_z=-20$~mag after including the samples from Fornax. It is driven by two ETGs, FCC~190 and FCC~249. Because they are not special in morphology, color, and location, we define $M_z=-20$~mag as the dividing line in this work, which is shown by the arrows. One possible reason why \citet{P08} did not find an increase of scatter at the brighter limit of $M_z\gtrsim-20$~mag is that the ACSVCS is incomplete for $B_T>12.15$~mag ($M_B>-18.94$~mag) and misses 63 ETGs from the parent sample. The typical $B-z$ color of massive ETGs is $\sim$2~mag, and the dividing line we choose in this work corresponds to $M_B\gtrsim-18$~mag, which is a traditional cut for defining early-type dwarf galaxies in observations. 

An outlier from this relation is FCC~47, which has $S_{N,z} = 2.51\pm0.11$ at $M_z=-20.14$~mag. It has the appearance of a normal ETG, and its color lies on the conventional ETG color-magnitude relation. However, it has the largest 3D clustercentric distance in the ACSFCS sample \citep{Blakeslee_09}, at $\sim2$~times of the virial radius of the Fornax Cluster.  From the projected galaxy distribution in Fornax, FCC~47 is relatively isolated.
One hypothesis is that it is an infalling central galaxy with a higher total mass-to-light ratio, resembling the behavior of the most massive ETGs.  Its GC system has possibly experienced fewer external disruption processes and the GCs may have a higher survival efficiency. 

The blue GCs are the dominant populations in most ETGs. The distribution of \snz\ is significantly biased by them. In the middle panel, the \snz\ of blue GCs mimics the distribution of \snz\ in the top panel. Low-mass ETGs have few to zero red GCs, something that is likely due to their low stellar metallicities. At higher masses, $S_{N,z,red}$ slightly increases with \Mz, on average, among both the intermediate- and the highest-mass ETGs. 

A more direct way to study the formation ability of red GCs is to look at their fraction of the total GC population. In the bottom panel, the red and blue crosses illustrate the relationship between \fRed\ and \Mz\ in the Fornax and Virgo Clusters. 
The low-mass ETGs that are fainter than $M_z=-20$~mag have relatively low \fRed\ scattered below 0.4. A fraction of the faintest ETGs do not have any red GCs. As with just the Virgo data, we find that \fRed\ is positively correlated with \Mz\ in the intermediate-mass range, and either flattens or decreases with \Mz\ at the high-mass end. 
Interestingly, among the massive ETGs, the galaxies with the highest \fRed\ are all in Fornax. 
In addition, the shape of the \fRed\ distribution shows that the rise in \fRed\ starts at $M_z=-20$~mag. 

If all GCs formed {\it in situ}, then \snz\ would be a simple expression of the GC formation efficiency in that galaxy. However, because the outer halos of massive ETGs are built by the accretion of low-mass satellites, the measured \snz\ is actually a combination of the GC formation efficiencies from all of the current galaxy's progenitors. If we want to measure the efficiency of {\it in situ} formation of the most massive progenitor, these GCs are likely to be at the galaxy's center. Typically, it is assumed that the {\it in situ} GC components are the red GCs, and the behavior of $S_{N,z,red}$ is described above, but we can also avoid imposing this interpretation and instead look only at the behavior of the innermost GCs. We therefore also measure \snz\ within the innermost 1\re\ to study the {\it in situ} GC formation of massive galaxies (Table~\ref{tabInOut}). For the massive galaxies in both ACSFCS and ACSVCS, we isolate their central 1\re\ area, count the GCs therein, and use the z-band luminosity of this area (i.e., half the total luminosity). The most massive ACSVCS galaxies have \re\ larger than the FOV, and we do not include them.

\begin{figure}
\epsscale{1.25}
\plotone{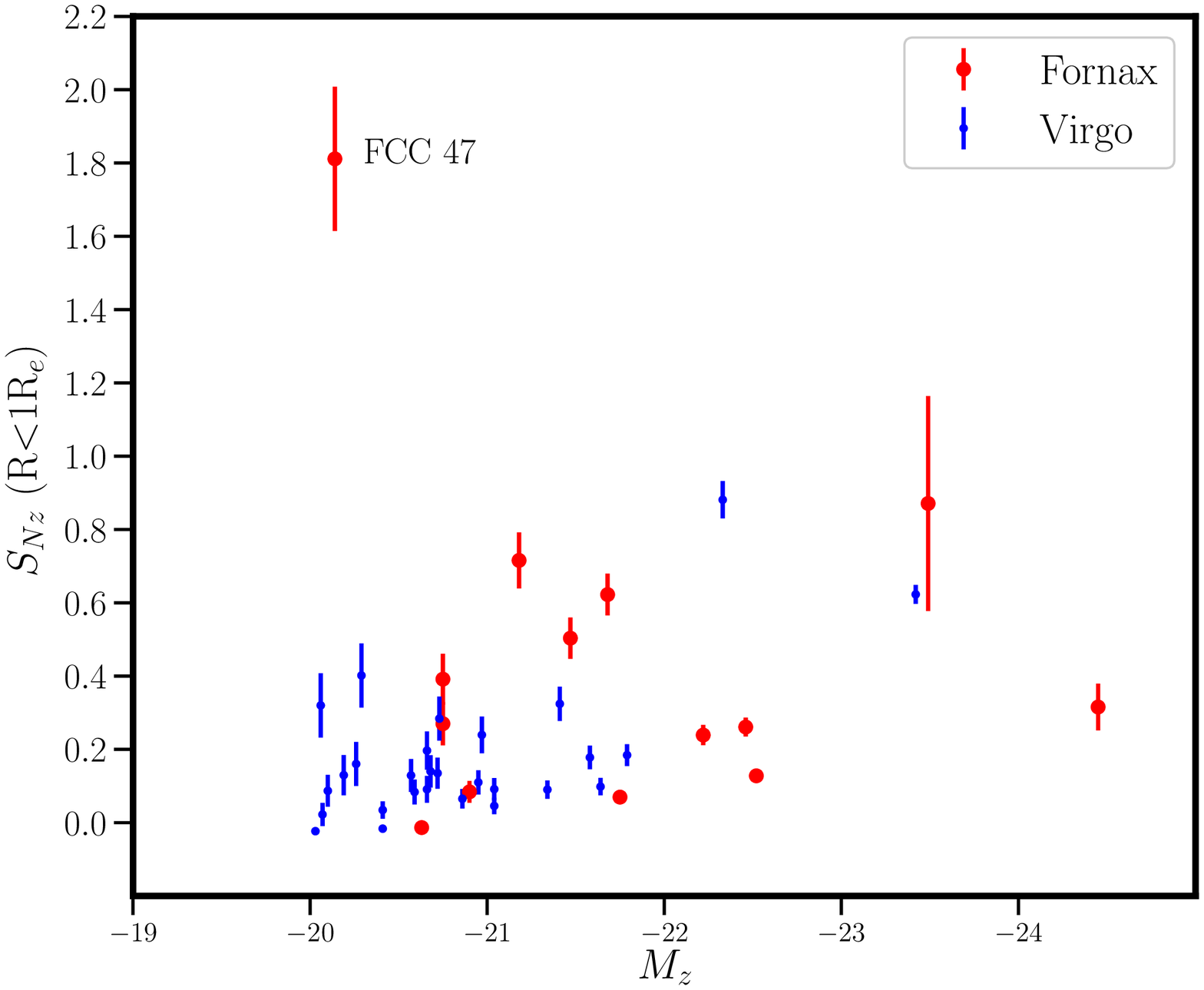}
\caption{ The \snz\ measured for the inner 1\re\ of massive ETGs in the ACSFCS (red circles) and ACSVCS (blue dots) samples and their relation to galactic $z$-band absolute magnitude. Excluding FCC~47, the outlier with a significantly high value, the brighter galaxies have higher inner \snz. \\
 \label{SN_iRe} }
\end{figure}

Figure~\ref{SN_iRe} shows the relation between this inner (in-situ) \snz\ and luminosities of massive galaxies. Excluding FCC~47, the outlier with a significantly high value, the brighter galaxies generally have higher inner \snz.

Regarding the outlier FCC~47, it has several special properties. It has the largest 3D clustercentric distance among ACSFCS sample. It is located at $\sim$2 times of the viral radius of the Fornax Cluster. Although it is relatively faint among massive galaxies, it has a rich GC system, which makes its \snz\ higher than others with similar masses. Furthermore, according to \citet{Liu_16b}, it has a significant sample of diffuse star clusters. However, we are not able to draw any conclusions for it with the limited information. 

% ------------- PLOTs --------------- %
% --------------------------------------- %

% ---------------------------------------- 
\subsection{Dependence on Environment}
\label{R_env}

\citet{P08} found that the low-mass ETGs at smaller clustercentric radii in Virgo have higher average \sn, indicating that the low-mass ETGs in the denser environments have higher GC formation efficiencies. In addition, they suggested that the environmental density is the second parameter that drives the large scatter of \snz\ at low masses. Correspondingly, simulation work by \citet{Mistani_16} found that dwarf galaxies in denser environments quenched earlier, boosting their total mass-to-light ratios and $S_N$.
In this section, we further explore the environmental effects on the GC formation in low-mass ETGs from the two clusters. 

In order to compare the environments in different clusters, we bring in two parameters to estimate the environment, following \citet{Guerou_15}. 
One is \SgmEnv, defined as the number of galaxies per square degree within a region that includes the 15 closest neighbors, indicating the number density of the local environment.  The other one is \LmL, the luminosity ratio between the brightest galaxy among the 15 closest neighbors and the galaxy itself. 
It generally reflects the mass ratio between the hosts and their neighbor galaxies.  In the Virgo Cluster, our calculation is based on the complete parent sample of 163 ETGs. Our \LmL\ is calculated in the $B$ band. Tables~\ref{tabEnv_F} and \ref{tabEnv_V} list the results of both parameters of ACSFCS and ACSVCS galaxies, respectively. 

Note that we have also tried using $\Sigma_{5}$ and \LmL, calculated with the five closest neighbors, and the results are generally the same. Both \SgmEnv\ and \LmL\ are calculated using projected distances. Although the line-of-sight distances of ACSFCS and ACSVCS galaxies are provided by \citet{Blakeslee_09}, we decided not to use them in our analysis because the relatively large error bars of the 3D distances would smear out the relations shown in the plots. The line-of-sight depth of the Fornax cluster is not resolved by the surface brightness distances used in \citet{Blakeslee_09}, so including these distances is of limited value.

% ------------- PLOTs --------------- %
\begin{figure}
\epsscale{1.25}
\plotone{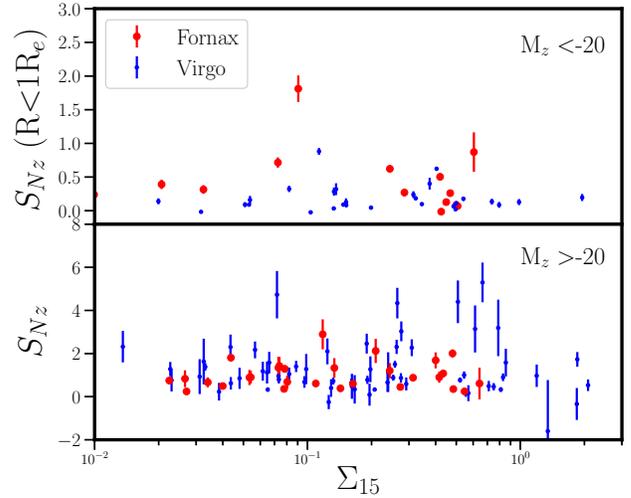}
\caption{ Upper: The \snz\ of the inner 1\re\ of massive ETGs and their relation with environmental density \SgmEnv. There is no clear environmental dependence of inner \snz. 
Lower: The relationship between the \snz\ of low-mass ETGs and \SgmEnv.  A fraction of \snz\ increases in very dense regions, but \snz\ is around zero in the densest environments. \\
 \label{SN_Sgm} }
\end{figure}
% ------------- PLOTs --------------- %
\begin{figure}
\epsscale{1.25}
\plotone{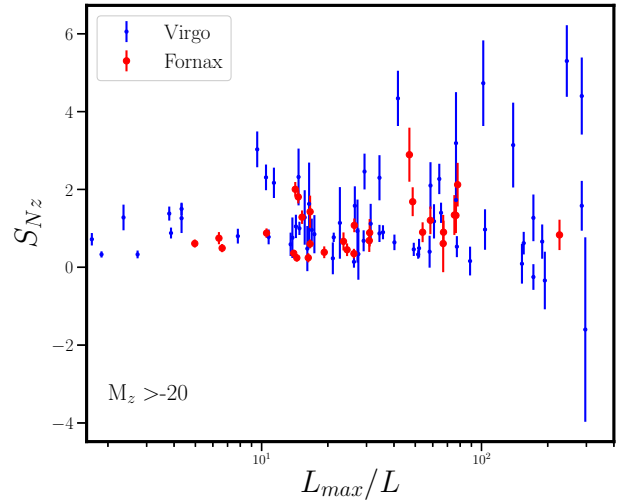}
\caption{ Relationship between the \snz\ of low-mass ETGs and \LmL\ in the $B$ band. The latter indicates the strength of the tidal field. Here \snz\ is roughly constant around 1.5 below \LmL$\sim10$. At higher \LmL, the scatter increases with \LmL. \\
 \label{SN_LmL} }
\end{figure}
% --------------------------------------- %

Figure~\ref{SN_Sgm} displays the relations between \snz\ and \SgmEnv. The relations for massive and low-mass ETGs are shown in the upper and lower panels, respectively.  In the upper panel, the measurements are limited to the inner 1\re, the same as in Figure~\ref{SN_iRe}.  There is no clear environmental dependence in this plot, implying that the in-situ formation of GCs in massive ETGs does not depend on the present-day environment (although environmental factors at higher redshift may have played a role).
In the lower panel, from a global view, the \snz\ values of low-mass ETGs are dispersed over the full range of \SgmEnv\ with a scatter larger than that of the intermediate-mass galaxies. The scatter significantly increases in very dense regions. In the very densest environments, however, all of the low-mass ETGs have \snz\ close to or consistent with zero. We propose that this is possibly due to tidal stripping from nearby massive galaxies. 

In Figure~\ref{SN_LmL}, we plot the \snz\ of low-mass ETGs against \LmL.  Below \LmL$\sim10$, \snz\ has a constant value around 1.5. At higher \LmL, the scatter increases with \LmL. These are galaxies that have a much more massive neighbor and indicate that the low-mass galaxies in denser environments could either have higher GC formation efficiencies or be significantly underpopulated with GCs (through inhibition or stripping). 

\begin{figure}
\epsscale{1.3}
\plotone{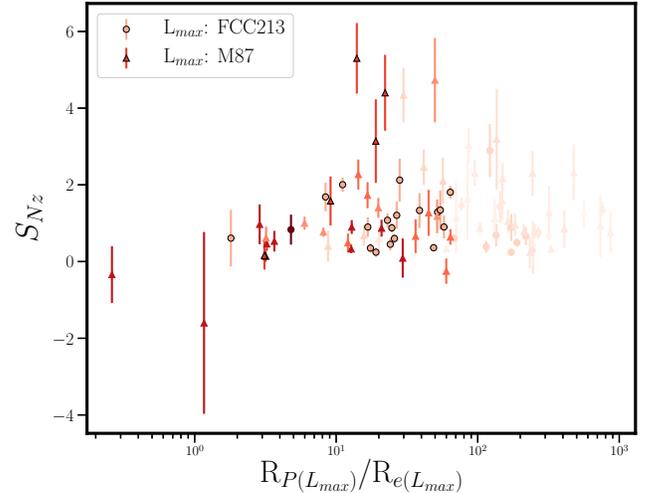}
\caption{Relationship between \snz\ of low-mass ETGs and their distance to the most luminous galaxy among the 15 closest neighbors (i.e., the galaxy with $L_{max}$). The distances are normalized by the \re\ of the galaxy with $L_{max}$. Circles and triangles represent galaxies from Fornax and Virgo, respectively. The colors represent the $L_{max}$, with the darker colors indicating higher $L_{max}$. The galaxies that have FCC~213 or M87 as their most luminous neighbors have black outlines. \\
 \label{SNz_rLm} }
\end{figure}

The strength of the tidal field a galaxy experiences is determined by not only the \LmL, but also the distance between galaxies.  In Figure~\ref{SNz_rLm}, we plot the \snz\ of low-mass ETGs against $R_{P(L_{max})}$, the projected distance from the most luminous galaxy among the 15 closest neighbors (i.e. the galaxy with $L_{max}$), and the distances are normalized by the \re\ of the galaxy with $L_{max}$. Circles and triangles represent galaxies from Fornax and Virgo, respectively. The colors indicate the luminosity of their most luminous neighbors (i.e. $L_{max}$), where the darker colors show the higher $L_{max}$. The galaxies that have FCC~213 or M87 as their most luminous neighbors have black outlines. 

All of the low-mass ETGs located closer than 10\re\ to their most luminous neighbors have low \snz. Similarly, all of the low-mass ETGs that have the most luminous galaxies in our sample as their most luminous neighbors have low \snz. Both suggest that the GC systems of these low-mass ETGs experienced tidal stripping. 
For the rest of the low-mass ETGs, at each distance, in general, the ones with the highest \snz\ have the most massive neighbors. 
In addition, their \snz, on average, decreases with the normalized distances. These indicate that apart from tidal stripping, a denser environment can also enhance the GC formation efficiency. 

% ------------- PLOTs --------------- %
% --------------------------------------- %

% =======================

\section{Discussion}
\label{discussion}

% ---------------------------------------- 
\subsection{Dependence on Galactic Halo Mass}
\label{D_mass}

The variation of GC specific frequency has been shown to follow that of the stellar mass-to-light ratio in galaxies---i.e., the kind of galaxies that have high $S_N$ (low- and high-mass) are also the kind that have high total $M/L$. This idea has been explored by associating galaxies with dark matter halos through abundance matching \citep[i.e.,][]{P08}, weak lensing \citep{Hudson_14}, or dynamical mass estimates \citep{Harris_13,Forbes_18}. 
% ------------- PLOTs --------------- %
\begin{figure}
\epsscale{1.25}
\plotone{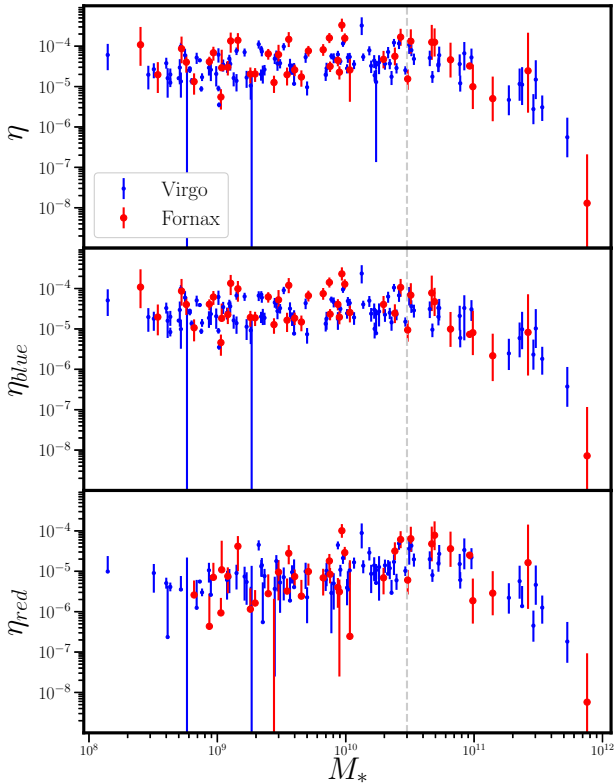}
\caption{ Galactic mass dependence of $\eta$ (the ratio of GC mass to galactic halo mass). The top, middle, and bottom panels show the $\eta$ distributions of the total, blue GC, and red GC populations, respectively. Red circles and blue dots represent the ETGs from Fornax and Virgo, respectively. For clarity, we do not plot the lower error bars, which are larger than the measurements themselves, but we use them in our analysis. 
In all panels, the GC mass fraction of ETGs more massive than $3\times10^{10}M_\odot$ stellar mass (dashed lines) decreases toward the high-mass end. Among the galaxies that are less massive, the $\eta$ values of blue and total GCs are generally constant, but that of red GCs slightly increases with mass. \\
 \label{eta_Mh} }
\end{figure}
% --------------------------------------- %

\citet{Hudson_14} studied GC mass fractions across a wide mass range. They found that the mass ratio of GCs to the halos of their host galaxies, $\eta$, is constant below a stellar mass $\mathcal{M}_\star=10^{11}M_\odot$ or a halo mass $\mathcal{M}_h=10^{13}M_\odot$, and slightly decreases toward the higher masses. 
However, the GC numbers of Fornax galaxies in their sample were adopted from \citet{Harris_13}, which are less accurate than what we present (see Section~\ref{Ngc}). Here we investigate this relationship with our improved measurements and more homogeneous dataset. 

We divide the GCs into blue and red populations in each galaxy and convert the total luminosity of these two populations into stellar masses by their mean colors using the 12~Gyr single stellar population model of \citet{BC03}. The total GC masses are the sum of these two populations. 
Note that the assumption of 12~Gyr is not suitable for FCC~21, which contains a substantial number of intermediate-age red GCs. To estimate the lower-mass limit, we estimate the total mass of its GC system by assuming that all of its red GCs have an age of 3~Gyr. In this case, the total GC mass in FCC~21 decreases by 25\%. Because the change is within the uncertainty, we do not show this lower-mass limit in the plot. 

Galactic halo masses are inferred from the galactic stellar masses according to the Equation~(21) in \citet{Behroozi_10} in the local universe. We use the scale factor for today (a=1), and adopt the best-fitting parameters at $\mu=\kappa=0$. Like in other work, these halo masses are thus not direct measurements but are inferred from a global stellar mass-halo mass relation.

Figure~\ref{eta_Mh} displays how $\eta$ distributes with galactic stellar mass. From top to bottom, the $\eta$ is calculated for the total GC population, blue GCs, and red GCs, respectively. Red circles and blue dots represent the ETGs from Fornax and Virgo, respectively. We only plot the data points with positive number of GCs but point out that the existence of galaxies with zero GCs (but, presumably, with a nonzero halo mass) shows that the relationship must break down at some point or has large scatter.

In all panels, the GC mass fraction decreases toward the high-mass end when M$_*$ exceeds $3\times10^{10}M_\odot$ (dashed lines), reproducing the trends in \citet{Hudson_14}. A possible reason is that the halo masses estimated by \citet{Behroozi_10} are high in this range. 
\citet{El-Badry_18} investigated this relation between GC masses and halo masses using a semi-analytic model built on dark matter merger trees. They produced a linear relation for massive galaxies, and the halo masses in their simulation are lower than what \citet{Behroozi_10} predicted at given stellar masses (see Figure~13 in \citealt{El-Badry_18}). It is also the case that the stellar mass-halo mass relation becomes very steep at these stellar masses, and so determining $\eta$ for these massive galaxies becomes problematic. 

Among the galaxies with lower masses, the $\eta$ values of blue and total GCs are generally constant. On the other hand, the mass fraction of red GCs increases slightly with mass, indicating a lower formation efficiency of metal-rich GCs in lower-mass halos. 
\citet{El-Badry_18} produced similar relations in their simulation. In addition, they found that the linear relation at the massive end is a result of the central limit theorem during the hierarchical assembly, independent of GC formation scenarios. However, for galaxies with lower masses, the physics of GC formation is required to reproduce the relations. 

If the GC formation efficiency with respect to halo masses is universal, as suggested by the constant mean $\eta$ at low masses, the intrinsic scatter of $\eta$ reflects the variation of field star formation efficiency, i.e., the scatter in $M_h/L$. 
For the low-mass ETGs with $M_z>-20$, we modify MPFITEXY \citep{Tremaine_02,Press_92}, allowing for fitting with asymmetric errors, and fit their zero-slope linear relation between $log(\eta)$ and $log(\mathcal{M}_\star)$. We perform 5000 bootstrap iterations on each fit to calculate the errors of our fitting parameters. The results show that they have a mean $\eta$ of $(3.36\pm0.25)\times10^{-5}$ with an intrinsic scatter of $0.30\pm0.04$~dex, in agreement with the results from other studies \citep{Georgiev_10,Hudson_14}. 

In our analysis, we do not include the uncertainty of the galactic stellar mass-halo mass relation. The simulations from \citet{Behroozi_10} showed a typical scatter of 0.15-0.2~dex, which is not able to fully account for the intrinsic scatter in our results. It implies that the GC mass fraction has scatter about 0.2~dex. Alternatively, the simulation underestimated the scatter of the $\mathcal{M}_\star-\mathcal{M}_h$ relation, and the variation of $\eta$ is smaller. \\

% ---------------------------------------- 
\subsection{GCs and the Origin of Nuclear Star Clusters} 
\label{D_nu}

The timescales for GCs to sink into the centers of low-mass ETGs by dynamical friction can be less than the Hubble time. \citet{Tremaine_75} suggested that galactic nuclei formed by the mergers of the GCs fallen into centers.  \citet{Lotz_01} examined this hypothesis by comparing the central deficiency of the GC radial density profiles and the brightness of their nuclei, and they found that the nuclei are brighter than expected. 
This could be a result of subsequent star formation in the galactic center, as \citet{Lotz_04} found that some nuclei in low-mass ETGs experienced recent star formation. It is also possible that some GCs had been disrupted into field stars during their inspiral, because the tidal force is increasing toward the galactic center. 

% ------------- PLOTs --------------- %
\begin{figure}
\epsscale{1.25}
\plotone{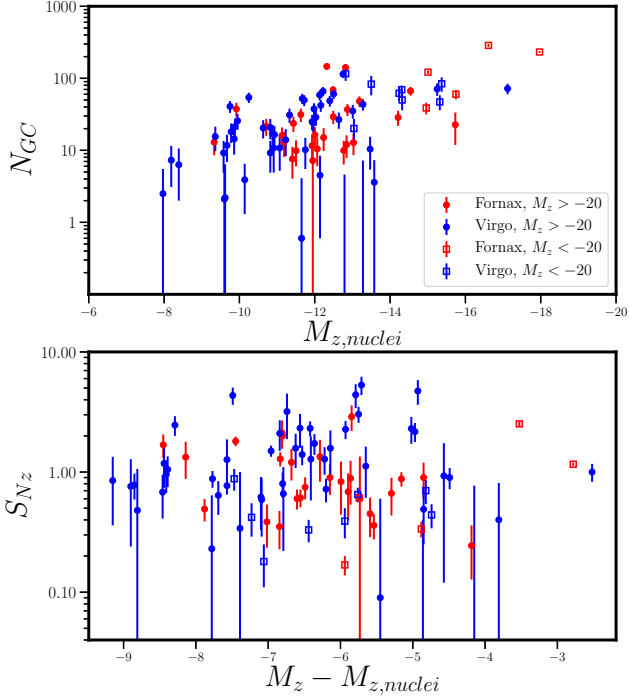}
\caption{ 
Upper: The relation between the number of GCs and the \Mz\ of the nuclei of their hosts among the nucleated ETGs. Red and blue symbols represent the galaxies from Fornax and Virgo. Open squares and filled circles represent massive and low-mass ETGs, respectively. The GC number is higher in the hosts with brighter nuclei. 
Lower: The relation between \snz\ and the magnitude difference between the nucleated galaxies and their nuclei, which are the normalization of the GC numbers and nuclear light by the luminosity of galactic field stars. These two parameters are not clearly correlated, indicating a weak link between the infall of GCs and nuclear formation. \\
 \label{SNz_Nu} }
\end{figure}
% --------------------------------------- %

Here we explore the link between GCs and nuclei, comparing the GC numbers with the properties of stellar nuclei. 
The upper panel of Figure~\ref{SNz_Nu} displays the relation between the number of GCs and the \Mz\ of the nuclei \citep{Cote_06,Blakeslee_09,Turner_12} in their hosts among the nucleated ETGs. Red and blue symbols represent the galaxies from Fornax and Virgo. Open squares and filled circles represent massive and low-mass ETGs, respectively. The GC number is higher when the hosts have brighter nuclei, which can be a result of two positive correlations. One is between the number of GCs and the mass of their host galaxies. The other is between the galactic luminosity and the nuclear luminosity \citep{Lotz_04,Cote_06}. 

However, when normalizing them to the galactic luminosity (lower panel of Figure~\ref{SNz_Nu}), \snz\ shows no dependence on the magnitude difference between the galaxies and their nuclei (i.e., the nuclear luminosity fraction).  If there is a constant fraction of stars that end up in GCs during early star formation, and then the nuclei are built up due to a fraction of GCs that sink into the center, we should expect an anticorrelation between \snz\ and $M_z-M_{z,nuclei}$ (i.e., larger nuclei means fewer remaining GCs). 
Even if a component of the nuclei is made of subsequent star formation, or some GCs are disrupted before sinking in, this anticorrelation might still be evident if the GC inspiral is the dominant mechanism. 
It appears that not all galactic nuclei are made from the GCs that spiraled into the centers. This is in agreement with the recent study of the stellar populations of galactic nuclei in the ACSVCS galaxies \citep{Spengler_17}. 

% =======================

\section{Summary}
\label{summary}

We measure GC specific frequencies (\sn, \snz) in 43 galaxies in the Fornax Cluster, a complete sample of Fornax ETGs brighter than $M_B\sim-16$~mag. Together with the homogeneous measurements of 100 ETGs in Virgo from \citet{P08}, we study the GC formation efficiency in ETGs over a wide range of galactic mass and environmental density. 

\begin{itemize}

\item The \snz\ of ETGs has similar properties in Fornax and Virgo. The one order of magnitude difference in the densities of these two galaxy clusters does not have a significant effect on GC formation in ETGs. 

\item The \snz\ of ETGs in both galaxy clusters have similar distributions with galactic z-band absolute magnitude, an indicator of galactic stellar mass. 
$S_{N,z}\sim 1.5$ for intermediate-mass ETGs and increases with \Mz\ at the bright end ($M_z<-23$~mag). In the low-mass range ($M_z>-20$~mag), it also rises on average, but with large scatter. 

\item The \snz\ of low-mass ETGs has an environmental dependence. 
(1) Low-mass ETGs that are located within 10\re\ of their massive host galaxies, or have the most luminous galaxies in our sample as their neighbors, have universally low \snz, showing that their GC systems have likely been tidal stripped. 
(2) For the remaining low-mass ETGs being closer to their massive neighbors (but not too close) results in them having higher \snz. At a given distance from their massive neighbors, the ones with higher \snz\ generally have more massive neighbors. These indicate that apart from tidal stripping, a denser environment can also enhance the GC formation efficiency. 

\item The mass ratio between GC systems and the halos of their host galaxies is constant with an intrinsic scatter of $\sim0.3$~dex when galactic stellar masses are below $3\times10^{10}M_\odot$ (corresponding to a halo mass of $3\times10^{12}M_\odot$). It slightly decreases towards the massive end in the higher mass range, where inferring halo masses from stellar masses becomes difficult. These are in agreement with previous studies. The constant GC mass fraction of low-mass ETGs ($\eta \sim3.36\times10^{-5}$ on average) suggests a uniform formation efficiency of GCs with respect to the dark matter halo masses, and suggests that the GC formation capability is fundamentally governed by the potential wells of galaxies. 

\end{itemize}

% =======================

\acknowledgements
Y.L. acknowledges support from the Oxford Centre for Astrophysical Surveys, which is funded through generous support from the Hintze Family Charitable Foundation. 
She also thanks Shuo Yuan and Long Wang for the helpful discussions on statistics. 
Y.L. and E.W.P. acknowledge support from the National Natural Science Foundation of China under grant No. 11573002, and from the Strategic Priority Research Program, ``The Emergence of Cosmological Structures'', of the Chinese Academy of Sciences, grant No. XDB09000105. 
T.H.P. acknowledges support through FONDECYT Regular project 1161817 and CONICYT project Basal AFB-170002.\\ 

% ------------ Tables ------------%
\begin{deluxetable}{lcc}
\tablewidth{0pt}
\tablecaption{ Environmental Parameters of ACSFCS Galaxies. 
 \label{tabEnv_F} }
\tablehead{
\colhead{ID} & 
\colhead{\LmL} & 
\colhead{\SgmEnv\ (L$_\odot$/pc$^2$)} 
}
\decimalcolnumbers
\startdata
%# galx  LB_max/LB  Sigma15(Lsun/pc^2)
FCC~21   & $ 0.093   $ & $ 0.032 $ \\ 
FCC~213  & $ 0.735   $ & $ 0.605 $ \\ 
FCC~219  & $ 1.359   $ & $ 0.448 $ \\ 
FCC~167  & $ 1.793   $ & $ 0.468 $ \\ 
FCC~276  & $ 2.710   $ & $ 0.243 $ \\ 
FCC~147  & $ 3.262   $ & $ 0.419 $ \\ 
FCC~83   & $ 2.520   $ & $ 0.072 $ \\ 
FCC~184  & $ 5.107   $ & $ 0.507 $ \\ 
FCC~63   & $ 3.747   $ & $ 0.020 $ \\ 
FCC~193  & $ 7.341   $ & $ 0.499 $ \\ 
FCC~170  & $ 8.227   $ & $ 0.424 $ \\ 
FCC~153  & $ 4.967   $ & $ 0.109 $ \\ 
FCC~177  & $ 10.531  $ & $ 0.313 $ \\ 
FCC~47   & $ 3.651   $ & $ 0.090 $ \\ 
FCC~43   & $ 6.614   $ & $ 0.039 $ \\ 
FCC~190  & $ 14.225  $ & $ 0.480 $ \\ 
FCC~310  & $ 13.993  $ & $ 0.077 $ \\ 
FCC~249  & $ 14.676  $ & $ 0.043 $ \\ 
FCC~148  & $ 16.296  $ & $ 0.547 $ \\ 
FCC~255  & $ 15.274  $ & $ 0.078 $ \\ 
FCC~277  & $ 16.616  $ & $ 0.163 $ \\ 
FCC~55   & $ 6.396   $ & $ 0.022 $ \\ 
FCC~152  & $ 14.456  $ & $ 0.027 $ \\ 
FCC~301  & $ 26.291  $ & $ 0.484 $ \\ 
FCC~335  & $ 24.470  $ & $ 0.272 $ \\ 
NGC~1340 & $ 0.985   $ & $ 0.009 $ \\ 
FCC~143  & $ 26.400  $ & $ 0.434 $ \\ 
FCC~95   & $ 19.269  $ & $ 0.142 $ \\ 
FCC~136  & $ 53.977  $ & $ 0.416 $ \\ 
FCC~182  & $ 48.650  $ & $ 0.400 $ \\ 
FCC~204  & $ 23.570  $ & $ 0.034 $ \\ 
FCC~119  & $ 31.005  $ & $ 0.053 $ \\ 
FCC~90   & $ 58.541  $ & $ 0.243 $ \\ 
FCC~26   & $ 16.616  $ & $ 0.073 $ \\ 
FCC~106  & $ 30.878  $ & $ 0.080 $ \\ 
FCC~19   & $ 226.324 $ & $ 0.026 $ \\ 
FCC~202  & $ 66.951  $ & $ 0.642 $ \\ 
FCC~324  & $ 76.500  $ & $ 0.133 $ \\ 
FCC~288  & $ 75.278  $ & $ 0.073 $ \\ 
FCC~303  & $ 67.181  $ & $ 0.054 $ \\ 
FCC~203  & $ 77.878  $ & $ 0.209 $ \\ 
FCC~100  & $ 46.994  $ & $ 0.117 $ \\ 
IC~2006  & $ 4.863   $ & $ 0.285 $ \\ 
\enddata
\tablecomments{ Both \LmL\ and \SgmEnv\ are calculated in the B band. }
\end{deluxetable}

\clearpage
\begin{longtable}{lcc}
\tablewidth{0pt}
\caption{ Environmental Parameters of ACSVCS Galaxies }
\label{tabEnv_V} 
\tabularnewline
\hline
ID & \LmL\ & \SgmEnv\ (L$_\odot$/pc$^2$) \\ 
\decimalcolnumbers
%# galx  LB_max/LB  Sigma15(Lsun/pc^2)
VCC~1226 & $ 0.304   $ & $ 1.826 $ \\ 
VCC~1316 & $ 0.091   $ & $ 0.589 $ \\ 
VCC~1978 & $ 0.310   $ & $ 0.404 $ \\ 
VCC~881  & $ 1.0     $ & $ 0.511 $ \\ 
VCC~798  & $ 0.161   $ & $ 0.044 $ \\ 
VCC~763  & $ 1.0     $ & $ 0.404 $ \\ 
VCC~731  & $ 1.629   $ & $ 0.453 $ \\ 
VCC~1535 & $ 3.280   $ & $ 2.116 $ \\ 
VCC~1903 & $ 3.221   $ & $ 0.422 $ \\ 
VCC~1632 & $ 0.319   $ & $ 0.113 $ \\ 
VCC~1231 & $ 3.162   $ & $ 0.540 $ \\ 
VCC~2095 & $ 3.564   $ & $ 0.500 $ \\ 
VCC~1154 & $ 1.158   $ & $ 0.322 $ \\ 
VCC~1062 & $ 0.586   $ & $ 0.053 $ \\ 
VCC~2092 & $ 5.495   $ & $ 0.344 $ \\ 
VCC~369  & $ 5.597   $ & $ 0.132 $ \\ 
VCC~759  & $ 0.258   $ & $ 0.082 $ \\ 
VCC~1692 & $ 10.185  $ & $ 0.198 $ \\ 
VCC~1030 & $ 5.199   $ & $ 0.820 $ \\ 
VCC~2000 & $ 9.462   $ & $ 1.951 $ \\ 
VCC~685  & $ 6.194   $ & $ 0.147 $ \\ 
VCC~1664 & $ 2.831   $ & $ 0.314 $ \\ 
VCC~654  & $ 6.426   $ & $ 0.152 $ \\ 
VCC~944  & $ 1.706   $ & $ 0.050 $ \\ 
VCC~1938 & $ 8.165   $ & $ 0.485 $ \\ 
VCC~1279 & $ 10.964  $ & $ 0.735 $ \\ 
VCC~1720 & $ 0.398   $ & $ 0.019 $ \\ 
VCC~355  & $ 1.202   $ & $ 0.031 $ \\ 
VCC~1619 & $ 5.105   $ & $ 0.151 $ \\ 
VCC~1883 & $ 21.478  $ & $ 0.133 $ \\ 
VCC~1242 & $ 3.837   $ & $ 0.375 $ \\ 
VCC~784  & $ 2.208   $ & $ 0.053 $ \\ 
VCC~1537 & $ 1.870   $ & $ 0.206 $ \\ 
VCC~778  & $ 2.831   $ & $ 0.103 $ \\ 
VCC~1321 & $ 2.728   $ & $ 0.065 $ \\ 
VCC~828  & $ 11.376  $ & $ 0.496 $ \\ 
VCC~1250 & $ 20.511  $ & $ 0.795 $ \\ 
VCC~1630 & $ 6.854   $ & $ 0.136 $ \\ 
VCC~1146 & $ 14.859  $ & $ 0.544 $ \\ 
VCC~1025 & $ 18.535  $ & $ 0.986 $ \\ 
VCC~1303 & $ 34.355  $ & $ 0.274 $ \\ 
VCC~1913 & $ 35.645  $ & $ 0.829 $ \\ 
VCC~1327 & $ 26.302  $ & $ 0.554 $ \\ 
VCC~1125 & $ 3.872   $ & $ 0.252 $ \\ 
VCC~1475 & $ 3.801   $ & $ 0.033 $ \\ 
VCC~1178 & $ 49.203  $ & $ 0.749 $ \\ 
VCC~1283 & $ 21.281  $ & $ 0.520 $ \\ 
VCC~1261 & $ 1.690   $ & $ 0.132 $ \\ 
VCC~698  & $ 4.325   $ & $ 0.257 $ \\ 
VCC~1422 & $ 7.798   $ & $ 0.073 $ \\ 
VCC~2048 & $ 40.179  $ & $ 0.160 $ \\ 
VCC~1871 & $ 51.999  $ & $ 0.710 $ \\ 
VCC~9    & $ 10.764  $ & $ 0.191 $ \\ 
VCC~575  & $ 51.522  $ & $ 0.809 $ \\ 
VCC~1910 & $ 64.268  $ & $ 0.309 $ \\ 
VCC~1049 & $ 103.752 $ & $ 1.193 $ \\ 
VCC~856  & $ 11.376  $ & $ 0.056 $ \\ 
VCC~140  & $ 14.321  $ & $ 0.082 $ \\ 
VCC~1355 & $ 13.551  $ & $ 0.291 $ \\ 
VCC~1087 & $ 10.471  $ & $ 0.262 $ \\ 
VCC~1297 & $ 88.715  $ & $ 0.570 $ \\ 
VCC~1861 & $ 76.559  $ & $ 1.852 $ \\ 
VCC~543  & $ 16.904  $ & $ 0.073 $ \\ 
VCC~1431 & $ 9.549   $ & $ 0.275 $ \\ 
VCC~1528 & $ 29.376  $ & $ 0.189 $ \\ 
VCC~1695 & $ 29.107  $ & $ 0.096 $ \\ 
VCC~1833 & $ 4.325   $ & $ 0.006 $ \\ 
VCC~437  & $ 65.463  $ & $ 0.088 $ \\ 
VCC~2019 & $ 2.355   $ & $ 0.022 $ \\ 
VCC~33   & $ 27.542  $ & $ 0.166 $ \\ 
VCC~200  & $ 60.813  $ & $ 0.061 $ \\ 
VCC~571  & $ 77.268  $ & $ 2.080 $ \\ 
VCC~21   & $ 14.723  $ & $ 0.013 $ \\ 
VCC~1488 & $ 17.378  $ & $ 0.047 $ \\ 
VCC~1779 & $ 21.086  $ & $ 0.038 $ \\ 
VCC~1895 & $ 13.803  $ & $ 0.022 $ \\ 
VCC~1499 & $ 139.315 $ & $ 0.612 $ \\ 
VCC~1545 & $ 41.686  $ & $ 0.264 $ \\ 
VCC~1192 & $ 194.088 $ & $ 1.841 $ \\ 
VCC~1857 & $ 15.703  $ & $ 0.098 $ \\ 
VCC~1075 & $ 26.546  $ & $ 0.066 $ \\ 
VCC~1948 & $ 16.143  $ & $ 0.161 $ \\ 
VCC~1627 & $ 58.076  $ & $ 0.129 $ \\ 
VCC~1440 & $ 34.355  $ & $ 0.043 $ \\ 
VCC~230  & $ 101.859 $ & $ 0.071 $ \\ 
VCC~2050 & $ 172.186 $ & $ 0.197 $ \\ 
VCC~1993 & $ 172.186 $ & $ 0.125 $ \\ 
VCC~751  & $ 155.596 $ & $ 0.043 $ \\ 
VCC~1828 & $ 58.613  $ & $ 0.124 $ \\ 
VCC~538  & $ 152.756 $ & $ 0.195 $ \\ 
VCC~1407 & $ 244.343 $ & $ 0.664 $ \\ 
VCC~1886 & $ 188.799 $ & $ 0.237 $ \\ 
VCC~1199 & $ 296.483 $ & $ 1.346 $ \\ 
VCC~1743 & $ 16.443  $ & $ 0.032 $ \\ 
VCC~1539 & $ 285.759 $ & $ 0.509 $ \\ 
VCC~1185 & $ 285.759 $ & $ 0.854 $ \\ 
VCC~1826 & $ 27.289  $ & $ 0.031 $ \\ 
VCC~1512 & $ 22.698  $ & $ 0.239 $ \\ 
VCC~1489 & $ 76.559  $ & $ 0.787 $ \\ 
VCC~1661 & $ 31.332  $ & $ 0.064 $ \\ 
\tabularnewline
\hline
\end{longtable}

% ---------------------------------%

%\clearpage
% =======================

%\bibliographystyle{apj}
\bibliographystyle{aasjournal}
\bibliography{ref_SN}

\clearpage 

% =======================
\end{document}